\documentclass[11pt]{amsart}
\usepackage{amssymb,amsmath,xcolor,graphicx,xspace,colortbl,textcomp, rotating}
\usepackage{amsfonts}
\usepackage{amsmath}
\usepackage{amssymb}
\usepackage{graphicx}
\usepackage{subfigure}
\usepackage{float}
\usepackage[font=small]{caption}
\usepackage{lipsum}

\newcommand\blfootnote[1]{%
  \begingroup
  \renewcommand\thefootnote{}\footnote{#1}%
  \addtocounter{footnote}{-1}%
  \endgroup
}
\graphicspath{{spirals-Jan5_graphics/}{spirals-Jan5_tcache/}{spirals-Jan5_gcache/}}
\DeclareGraphicsExtensions{.pdf,.eps,.ps,.png,.jpg,.jpeg}
\providecommand{\U}[1]{\protect\rule{.1in}{.1in}}
\pdfoutput=1

\theoremstyle{plain}

\numberwithin{equation}{section}
\input{tcilatex}

\begin{document}
\title{On the initiation of spiral waves in excitable media}
\author{S. P. Hastings\textsuperscript{1} and M. M. Sussman\textsuperscript{2}}

\blfootnote{\textsuperscript{1} Department of Mathematics, University of Pittsburgh, sph@pitt.edu,  \textsuperscript{2} Independent Researcher, 5026 Belmont Ave, 
Bethel Park, PA 15102, mmsussman@gmail.com }

\begin{abstract}
Excitable media are systems which are at rest in the absence of external
input but which respond to a sufficiently strong stimulus by sending a wave
of \textquotedblleft excitation\textquotedblright\ across the medium.
Examples include cardiac, cortical, and other tissue in the central nervous
system, and in each of these examples, evidence of
rotating spiral waves has been observed and associated with abnormalities. 
How such waves are initiated is not well understood. In this numerical study
of a standard mathematical model of excitable media, we obtain spirals and
other oscillatory patterns by a method, simple in design, which had
previously been ruled out. We analyze the early stages of this process and
show that long term stable oscillatory behavior, including spiral waves, can
start with very simple initial conditions, such as two small spots of
excitation and all other areas at rest, and no subsequent input or disturbance. 
Thus, there are no refractory cells in the initial condition.  
For this model, even random spots of stimulation result in 
periodic rotating patterns relatively often, 
leading us to suggest that this could happen in living tissue.
\end{abstract}

\title{On the initiation of spiral waves in excitable media}

\maketitle
\section{Introduction}

\label{intro}
Spiral and other repetitive waves in excitable media are associated with
cardiac abnormalities \cite{Wiener, Davidenko, Panilov}, cortical oscillations \cite%
{Troy}, retinal spreading depression \cite{Gorelova}, traveling waves of
spreading depression in the central nervous system \cite{Mair}, 
spiral waves in the epileptic
neocortex \cite{Mair2}, patterns in the
Belousov-Zhabotinsky reaction \cite{Muller1}, aggregations of Dictyostelium
discoideum amoebae \cite{Gerisch, Sawai}, and other systems \cite{Zykov1}.
There is good theoretical work on fully formed waves, early references
including \cite{Keener, KeenerTyson}, but how they are initiated is not 
well understood \cite{Zykov1}%
. Mechanisms which have been proposed depend either on breaking an existing wave \cite{Zykov1},
stimulations at two
separate times, \cite{Winfree, Zykov1}, the medium being inhomogeneous 
\cite{ZykovBodenschatz, Zykov1}, or there being internal barriers with
corners \protect\cite{keenerMuller, Zykov1}. Here we present the results of
numerical computations on a standard partial differential equation (PDE)
system model for such studies, the
FitzHugh-Nagumo equations \cite{FitzHugh, Nagumo1, Nagumo2}, which indicate
that repetitive oscillations, including spirals, can be generated by an
initial multi-point stimulation in the interior of a spatially homogeneous
simply connected region and no further stimulation or disturbance.\ This
possibility was ruled out over 70 years ago \cite{Wiener}, based on physical
hypotheses later found to be in part unrealistic \ \protect\cite{Keener, Muller2,
Muller3}. However to our knowledge, no one has challenged the result in \cite%
{Wiener}\ in print since. We have found simple stimulation patterns which
lead to oscillatory behavior and which we hope are accessible in the
laboratory, and we have developed some qualitative understanding of the
underlying physical processes which occur in several of our examples.

In all of our examples these patterns involve waves with
the shape of an incomplete or broken ring that we term $C$ shaped.
These can lead to repeated activity, as we shall see. 

\begin{center}
\begin{figure}[]
\begin{tabular}{ccc}
\includegraphics[width=1.4 in, height=1.4 in] {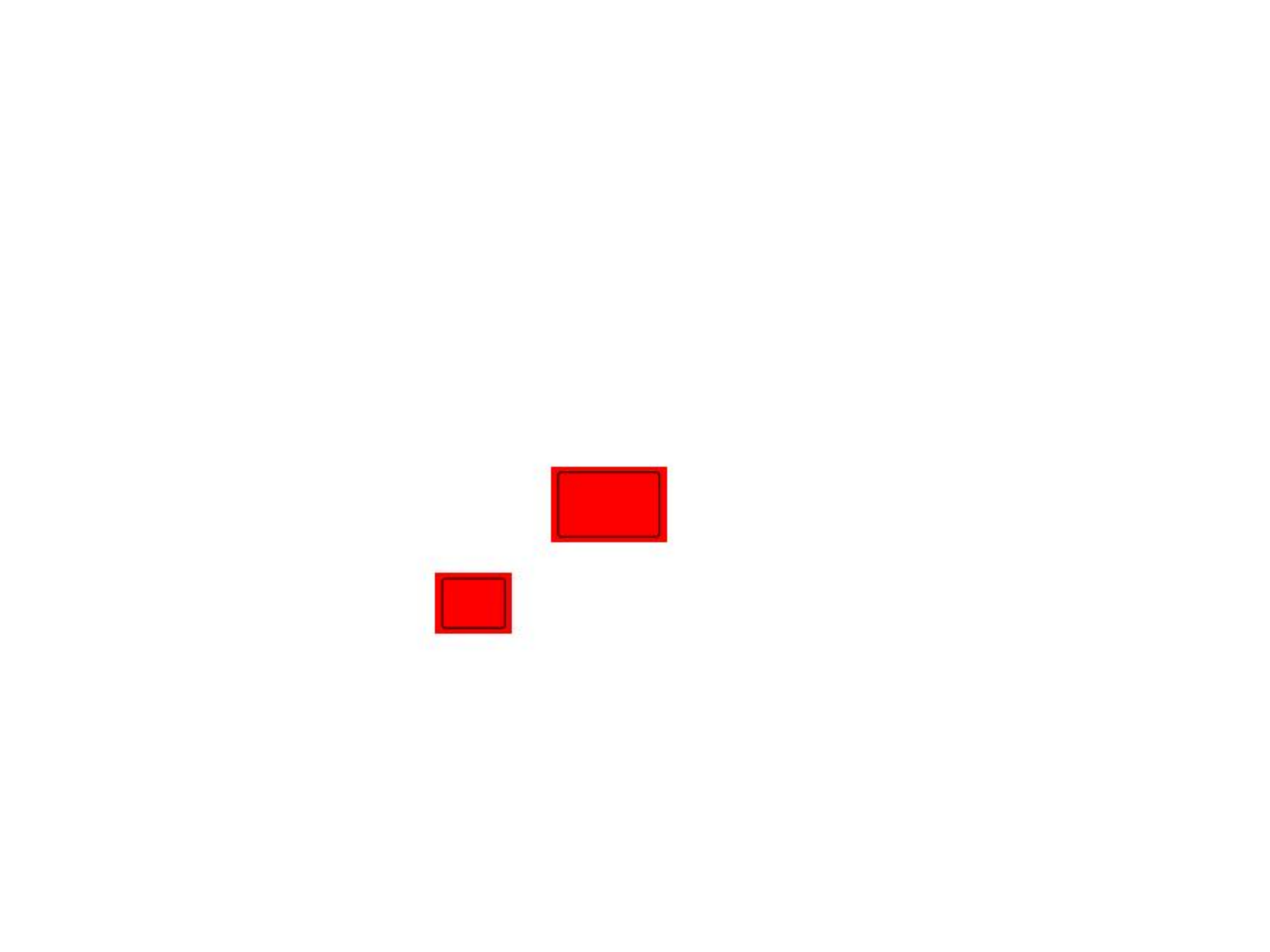} & %
\includegraphics[width=1.4 in, height=1.4 in] {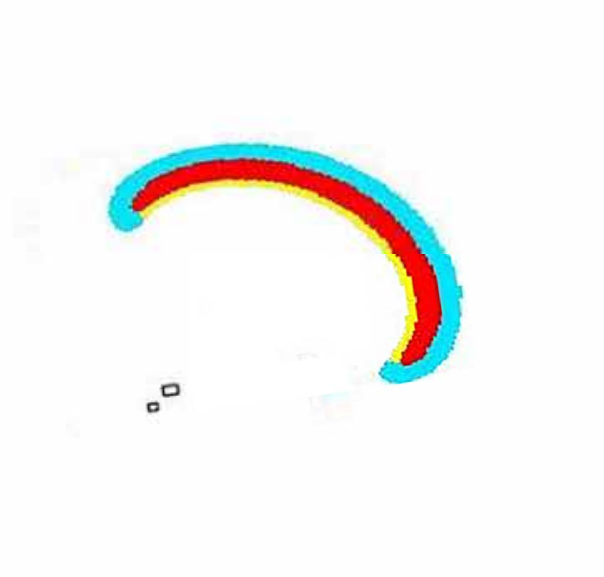} \\ 
(a) $t=0$ & (b) $t=755$ \\ 
&  
\end{tabular}%
\caption{This shows the principal discovery of this work, namely that a simple
pattern of initial excitation in an excitable medium, 
such as the pattern in (a) of two excited spots and all else at rest, can lead to an outgoing C-shaped wave, 
as in (b), with no external intervention. (It is known that C-shaped waves can lead to repetitive
behavior, as discussed in detail later in this paper.) In (b) only the region of depolarization is shown, but
a hyperpolarized region is 
also present at the time $t=755$ of (b) and it is   
included in figures below and
in a movie at https://pitt.box.com/s/qab0a69bynx78qx3jso7mts4rbqeq9t8, which gives a good idea of the development to this point. 
The scale change between (a) and (b), is shown by the two small ``marker'' spots behind the wave which 
are outlines of the spots in (a). 
The color code
 is given in the caption to Figure \ref{phaseplane1}. }
 
\label{twospotsinitial}
\end{figure}
\end{center}

Figure \ref{twospotsinitial} briefly illustrates a very simple example.  We will fill in
details below, after an introduction to the terminology we will use in the subsequent
descriptions of what we believe are important features seen in our numerical experiments. 
We urge the reader to view the accompanying movies as they are mentioned, to bring the figures to life.

 Further along in 
the text there is a detailed discussion of the major steps between (a) and (b), and even further along, 
a link to a movie carrying the pattern progression in this example to the point where it 
has become repetitive.

In the sequel, after introducing the mathematical model, we present another
example of an initial configuration that gives rise to repetitive behavior
starting from a small excited region in a field otherwise at equilibrium.
As in the example in Figure \ref{twospotsinitial}, no disturbance is required
after the initial excitation. Subsequently, we present a discussion of
a somewhat more complicated case of repetitive behavior, starting from
randomly generated excited regions and, again, without further
excitation.  We close with short discussions of two cases that give
rise to more elaborate patterns involving spirals.

\section{Mathematical models of excitable media}

\label{models}

Excitable media are characterized by having three identifiable
states,\textquotedblleft rest\textquotedblright , \textquotedblleft
excited\textquotedblright , and \textquotedblleft
refractory\textquotedblright\ and by not supporting spatially homogeneous
oscillations.\ The rest, or equilibrium, state is stable to small
perturbations. \ A \textquotedblleft cell\textquotedblright , or region, in
the excited state influences neighboring cells or regions which are in the
rest state by diffusion. \ The strength of the diffusion is sometimes 
sufficient to
move a neighboring rest state region into the excited state. \ An excited
state soon becomes ``refractory'', a state in which the cell neither
stimulates, nor is stimulated by, its neighbors. Refractory regions
eventually return to rest, but the time spent in the refractory state is
usually significantly longer than the time that a region is excited.

In the caption to Figure \ref{twospotsinitial} we used older terms from neuroscience,
``depolarization'' and ``hyperpolarization'', because we believe that they are widely
understood and we haven't yet introduced the FitzHugh-Nagumo model, which allows more precision.
In the caption to Figure \ref{phaseplane1} we will clarify the relation between
these two terms and the particular definitions of states in this model which we use from then on.

The simplest type of mathematical model for an excitable medium is a
cellular automaton (CA) of the type described in \cite{GH}, and more
generally in \cite{GGH}. It turns out that the models in \cite{GGH} are too
simple to support the behavior described in the first paragraph above, so we
will not discuss them here.

The model in \cite{Wiener},  is sometimes described as a CA, 
but careful
reading reveals that it is more sophisticated, with continuous time, space,
and state. The discussion is geometric, with waves of excitation moving
around the homogeneous medium according to certain rules. From these rules it
was deduced as a \textquotedblleft theorem\textquotedblright\ that no wave
exhibiting \textquotedblleft flutter\textquotedblright\ could be initiated
with just regions of rest and of excitation. \ By flutter the authors meant
\textquotedblleft self-perpetuating steady-state waves moving around the
closed path in one direction.\textquotedblright\ Rotating spiral waves are
one example.

One of the rules in \cite{Wiener} is: \textquotedblleft cardiac impulses,
once started, spread with a constant velocity equal in all directions as far
as the network continues''. \ It is this hypothesis which was later shown to
be non-physical. And it is crucial in the proof of the theorem in \cite%
{Wiener}. But \cite{Wiener} predated the introduction by Turing \cite{Turing}
of partial differential equations as viable models for such systems. \ Since
then, studies of such models and also experiments have made it apparent that
the speed of advance of a wave front in an excitable medium at any given
point depends on the curvature of the wave at that point. \ Perhaps the
first report of this was in \cite{Winfree}. In \cite{Keener} and \cite{KeenerTyson1}, asymptotic
methods are used to show the effect, and some early experiments are in \cite%
{Muller1}. However both the theory and the experiments studied existing
fully formed rotating waves. To this day, experiments generally start with a
stimulus that produces an outgoing wave. \ Once it is fully developed, some
method is used to break the wave front, which allows the ends of the wave to
circle around and eventually restimulate the initial area of the wave break.
In other cases, inhomogeneous media, or special types of interior obstacles,
create the same effect.

Recently we found that if the very simple cellular automaton in \cite{GH} is
modified in one simple way, continued activity can occur from an initial
condition containing only excited and rest cells and with no subsequent
disturbance. \ This activity seemed completely unphysical, more like \textquotedblleft chaos"
than spirals, and so we will not give details here. But with this motivation,
we began our study of the FitzHugh-Nagumo equations. 

\subsection{The numerical model}

The FitzHugh-Nagumo model can be expressed as the system of equations
\label{numerical model}
\begin{align}
u_{t}& =D\Delta u+au\left( 1-u\right) \left( u-b\right) -v  \label{2.1} \\
v_{t}& =\varepsilon (cu-v).  \label{2.2}
\end{align}
In these equations, using terms from neuroscience, $u$ represents the membrane potential in a nerve cell and $v$ represents a recovery variable, such as the conductance of the membrane. While the system of equations (\ref{2.1}) - (\ref{2.2})  has no quantitative 
relation to a particular physical system, its behavior is similar to that of many models from neuroscience and other 
areas mentioned in the first paragraph above. 
Throughout our study,
$D=10^{-5},a=0.5,c=0.5,\varepsilon =.002$,
while $b$ will be chosen at or close to $0.17$ in all our examples.

We used Neumann boundary conditions in all cases except a few tests 
which showed that Dirichlet boundary conditions gave similar results.  See
the section on discretization below. 

Our numerical calculations suggest that with our other parameters as given
above, if the \textquotedblleft threshold" $b$ is $0.2$ or higher then all
solutions tend to rest. On the other hand. for $b= 0.1,$ spirals had been
seen previously, but with refractory cells present from the beginning \cite%
{Weimar}. We saw long term persistence of excited cells when $b$ was in the
range from $0.165$ to $0.19$. We did not try to find the maximal interval in 
$b$ which supported this behavior.\footnote{When $b$ is close to where spirals 
cease to exist, (other parameters being held constant), the system is said to
be ``weakly excitable''.}

Our computations were based on square meshes of different sizes, say $%
M\times M$. We will denote the region covered by such a mesh by $\Omega$.
The spatial discretization was based on \textquotedblleft
cells\textquotedblright\ of size $\Delta x=\Delta y=.005.$\footnote{%
Further details of our numerical methods are given in the section on
discretization.} Thus, as $M$ changed, so did the size of the domain $%
\Omega$. We chose $u\left( x,y,0\right) $ in each cell to be either $0$ or $%
0.8,$ the latter easily being shown to be in the excited\ range for this
model with our parameters, and we set $v\left( x,y,0\right) =0$ for all $%
\left( x,y\right) $.

\subsection{The phase plane}

To describe how waves develop and propagate, we have added two new
states to the traditional
list of \textquotedblleft resting\textquotedblright , \textquotedblleft
excited\textquotedblright , and \textquotedblleft
refractory\textquotedblright , which we call ``wave front'' and ``wave
back''. These bridge the gap in the $u$ variable between the excited and the
rest and refractory states. More information is in the caption to Figure \ref%
{phaseplane1}, and in our analysis below of the sequence of images in
this figure.

Many authors have discussed the fronts and backs of waves in the context of
excitable media; see \cite{Wiener, Zykov1} for example. \ Usually, however,
one or both of these are thought of as curves, separating the excited region
from other states. Because our model is continuous, indeed smooth, it makes sense for both
our wave front and wave back to have significant positive area. We will see that
behavior in these regions is crucial for the development of spirals and other repetitive
patterns. 

\begin{figure}[]
\begin{center}
\includegraphics[height=1.9 in, width =2 in]{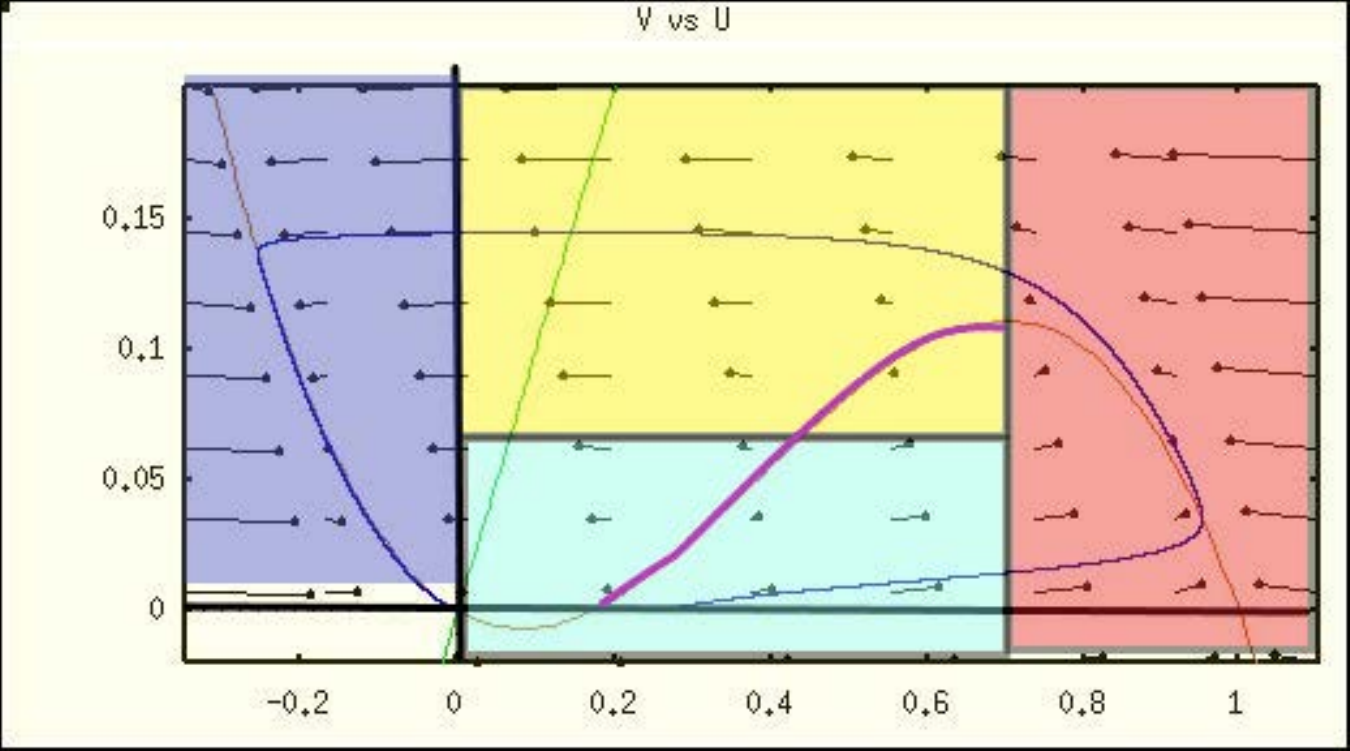}
\end{center}
\caption{The phase plane for the system (\ref{2.3},\ref{2.4}) of odes in variables 
$U,V$ obtained 
by dropping diffusion from the model. The $U$ and $V$ nullclines are shown, 
with the important middle branch of the $U$ nullcline in magenta. Color Code for
different states: red=excited ($U\geq 0.7$); dark blue=refractory, ($%
U\leq 0,$ $V>0.01$ ) light blue = wave front, ($0<U$ $<0.7,$ $V<0.06),$
yellow= wave back ($0<U<0.7,$ $V\geq 0.06$), all else is white, referred to
as the rest state. (Referring to the terms used in the caption to Figure \ref{twospotsinitial},
the depolarized region is the set $U>0$ and the hyperpolarized region is the set $U<0$.)  
A trajectory is shown proceeding in the ``standard order'' through the
wave front, excited, wave back, and refractory states and back to rest. }
\label{phaseplane1}
\end{figure}

We will say that a ``wave'' has formed when the three states ``excited'', ``wave front'', and ``wave back'',
as defined in the caption to Figure \ref{phaseplane1}, all exist and form locally a structure of relatively constant 
shape and moving in a particular direction, with the wave front followed by the excited region and then
the wave back.  However in the initial stage of
pattern development, when indeed the words ``wave'', ``front'', and ``back''
are somewhat misleading, this is not the case. In our analysis of our
graphical results, at those stages of the development when there are not yet
waves or wave fronts or backs, we will simply refer to these regions by
their color.  

As the pattern evolves, both the diffusion term $%
D\Delta u$ in (\ref{2.1}) and the other terms in both equations, referred to as either the
\textquotedblleft dynamics\textquotedblright\ or \textquotedblleft
reaction\textquotedblright\ terms, are important, but at different times and
places. A more refined graphical
presentation results from plotting contour lines for the $u$ and $v$
variables, and we will give an example of this.

In order to explain the very earliest steps of our examples, we first
give a brief introduction to the phase plane for the system of ordinary
differential equations obtained by dropping the diffusion term in (\ref{2.1}%
). Using variables $U,V$ to distinguish the solutions of this ode system
from solutions $(u,v)$ of (\ref{2.1})-(\ref{2.2}), the phase plane equations are  

\begin{align}
U^{\prime }& =aU\left( 1-U\right) \left( U-b\right) -V  \label{2.3} \\
V^{\prime }& =\varepsilon (cU-V)  \label{2.4}
\end{align}

Figure \ref{phaseplane1} shows the phase plane when $a=0.5,$ $b=0.17,$ $c=0.5,$
and $\varepsilon =.002.$
In that figure, let $\gamma $ denote the middle branch of the $U$
nullcline. Since for each $V$, $U^{\prime }<0$ in the region from $U=0$ to $%
\gamma $ and $U^{\prime }>0$ in the region from $\gamma $ to the right
decreasing branch of this nullcline, points on $\gamma $ are unstable
equilibrium points of (\ref{2.3}) when $V$ is taken as a constant. In Figure %
\ref{phaseplane1} we also show a trajectory starting at $(0.3,0)$, a point
somewhat to the right of $\gamma $. \ Note that it passes through the
colored regions in what we will call the \textquotedblleft standard
order\textquotedblright\ of states or colors:  
\begin{center}
$\text{wave front}\rightarrow\text{excited}\rightarrow\text{wave back}%
\rightarrow\text{refractory}\rightarrow \text{rest,}$ 
\end{center}
or 
\begin{center}
$\text{light blue}\rightarrow\text{red}\rightarrow\text{yellow}%
\rightarrow\text{dark blue}\rightarrow\text{white,}$
\end{center}
and never again leaves the rest state, since $(0,0)$ is an asymptotically
stable equilibrium point for (\ref{2.3},\ref{2.4}).\  (In this diffusionless 
setting there are no waves, wave backs, or wave fronts, so it is perhaps better 
to think of the colors only as identifying different regions of the plane for
later reference.)\ 

\begin{figure}[ht]
\begin{center}
\begin{tabular}{cc}
\includegraphics[height=1.7 in, width =1.7 in]{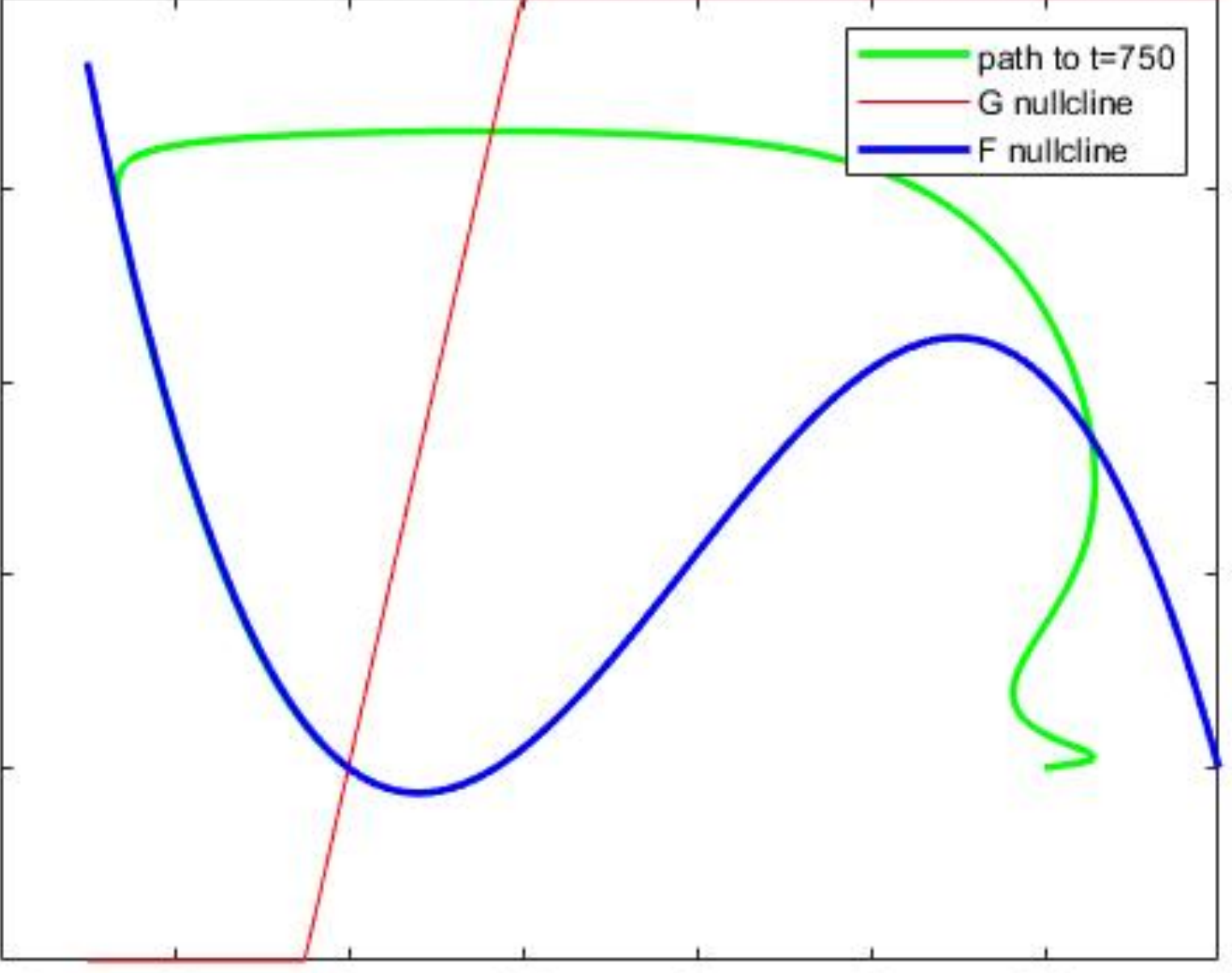} & %
\includegraphics[height=1.7 in, width =1.7 in]{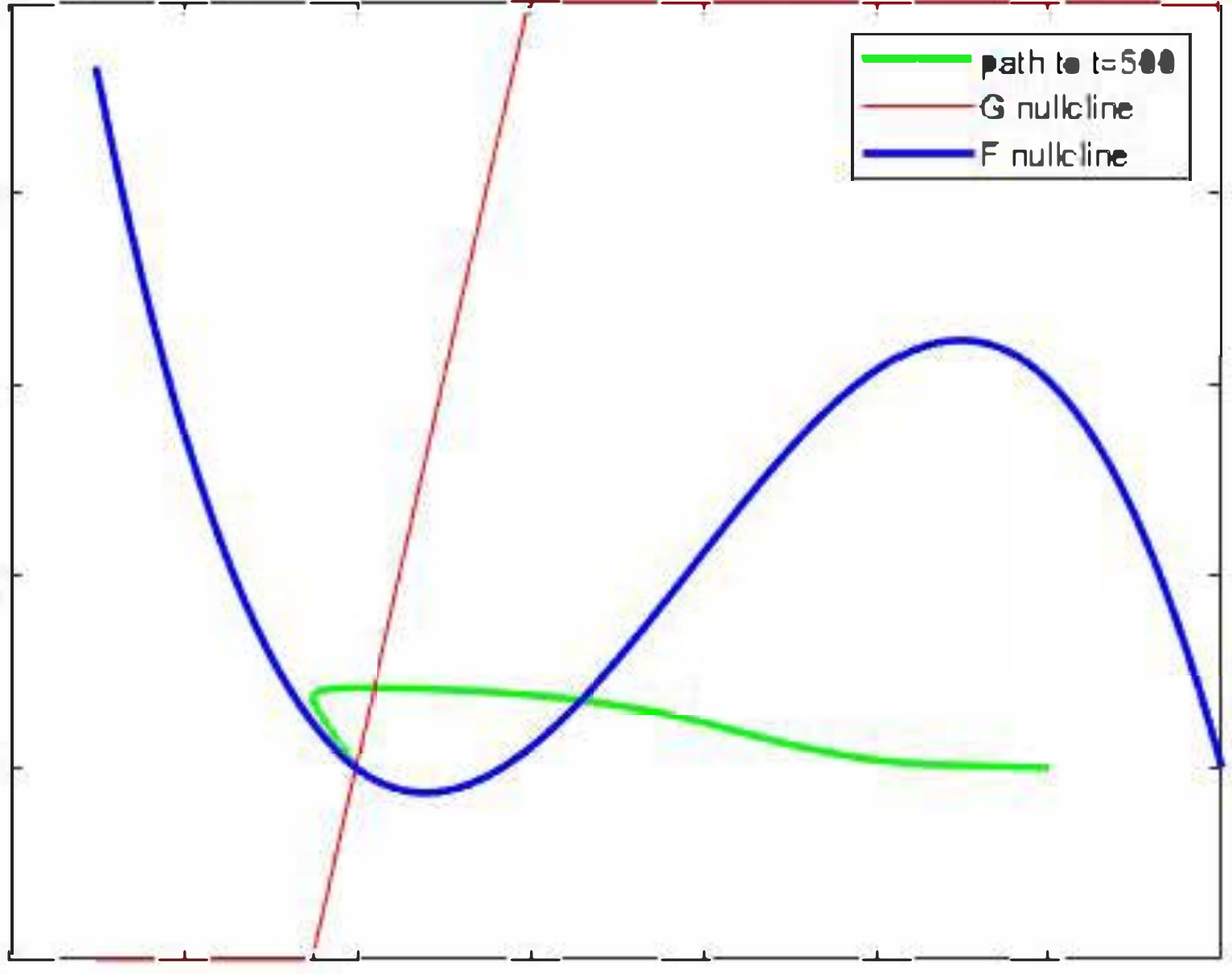} \\ 
(a) large initial spot & (b) small initial spot
\end{tabular}%
\end{center}
\caption{Green curves are plots of curves $t\rightarrow(u(x,y,t),v(x,y,t))$ for two solutions of 
the full model, with diffusion. (We call these ``PS curves''. Compare with Figure \ref{phaseplane1} .) In (a) the starting 
configuration was a single $6$ cell $\times 6$ cell spot 
of excitation
in a field otherwise at rest, and (x,y) was in the center of the spot. (b) was similar 
except the spot was $4\times 5$.}
\label{phaseplane2}
\end{figure}

In Figure \ref{phaseplane2} we consider plots of curves of the form $%
t\rightarrow (u\left( x,y,t\right) ,v\left( x,y,t)\right) $ \ for some
specific points $\left( x,y\right) $ in our domain, where $(u,v)$ is a
solution to (\ref{2.1}),(\ref{2.2}). We will refer to any curve of this sort
as a pseudo-phase plane, or \textquotedblleft PS\textquotedblright , curve. 
If a fully
developed wave crosses the point $(x,y)$ away from any end point this wave
may have, then the corresponding part of the PS curve passes through the
colors in the standard order. However this is not always true near a wave
tip or for the seed of a developing wave.

In Figure \ref{phaseplane2} we use PS curves to follow the development of two isolated spots,
one larger and one smaller. This figure refers to two separate simulations,
in each of which one of these spots is the only initially stimulated region in $%
\Omega$, all else starting at equilibrium. The larger spot generates an
outgoing target wave which reaches the boundary of any domain $\Omega$ in which it is placed. 
The smaller spot, if placed in the center of a sufficiently large domain $\Omega$, decays to rest without
generating a wave reaching the boundary. 

At the center of the large rectangle,
$u$ decreases into the blue region due to diffusion from below, but
then the positive and negative diffusive effects from areas of higher or
lower $u$ become more balanced, allowing the reaction terms to pull this
area back to the excited state.
From there, most cells in the larger rectangle follow the
normal order of states back to rest, though they may not remain
there. In contrast, all points in the smaller rectangle are subject to greater
diffusive effects from the regions where $u$ is small than from nearby
excited cells, which causes $(u,v)$ on the PS curve to cross the $U$
nullcline. From that point, reaction and diffusion from rest areas are
acting in the same direction, with reaction now dominating and pushing $u$
negative, and so $(u,v)$ becomes refractory without ever entering the yellow
region. This is not, therefore, the standard color order.

The small rectangle example shows how a refractory region can be produced in
such a way that it is not shielded behind an advancing wave, and so can
interact with something coming toward it. The timing of these interactions is also crucial 
and should be the object of future research. 

We will call a rectangular spot of excited cells which, when placed in a field otherwise at rest, 
does not generate any
wave which reaches the boundary of a sufficiently large domain $\Omega$
``subcritical''. This is a property which depends only on the dimensions $m$
and $n$ of the spot.\footnote{%
We believe, based on our simulations, that for any given $m$ and $n$ there
is an $M$ such that if an $m \times n$ spot of excitation is entirely contained in a
square of side $M/2$ centrally located within the $M \times M$ domain $%
\Omega $, and it generates a wave which reaches the boundary of $\Omega$,
then this wave will reach the boundary of any larger square domain
containing $\Omega$. An $M$ with this property therefore will be
``sufficiently large'' in our definition of ``subcritical''.}

\section{A simple model giving rise to repetitive motion \label{2spots}}

\begin{figure}[]
\begin{center}
\begin{tabular}{cc}
\includegraphics[height=1.0 in, width =1.0 in]{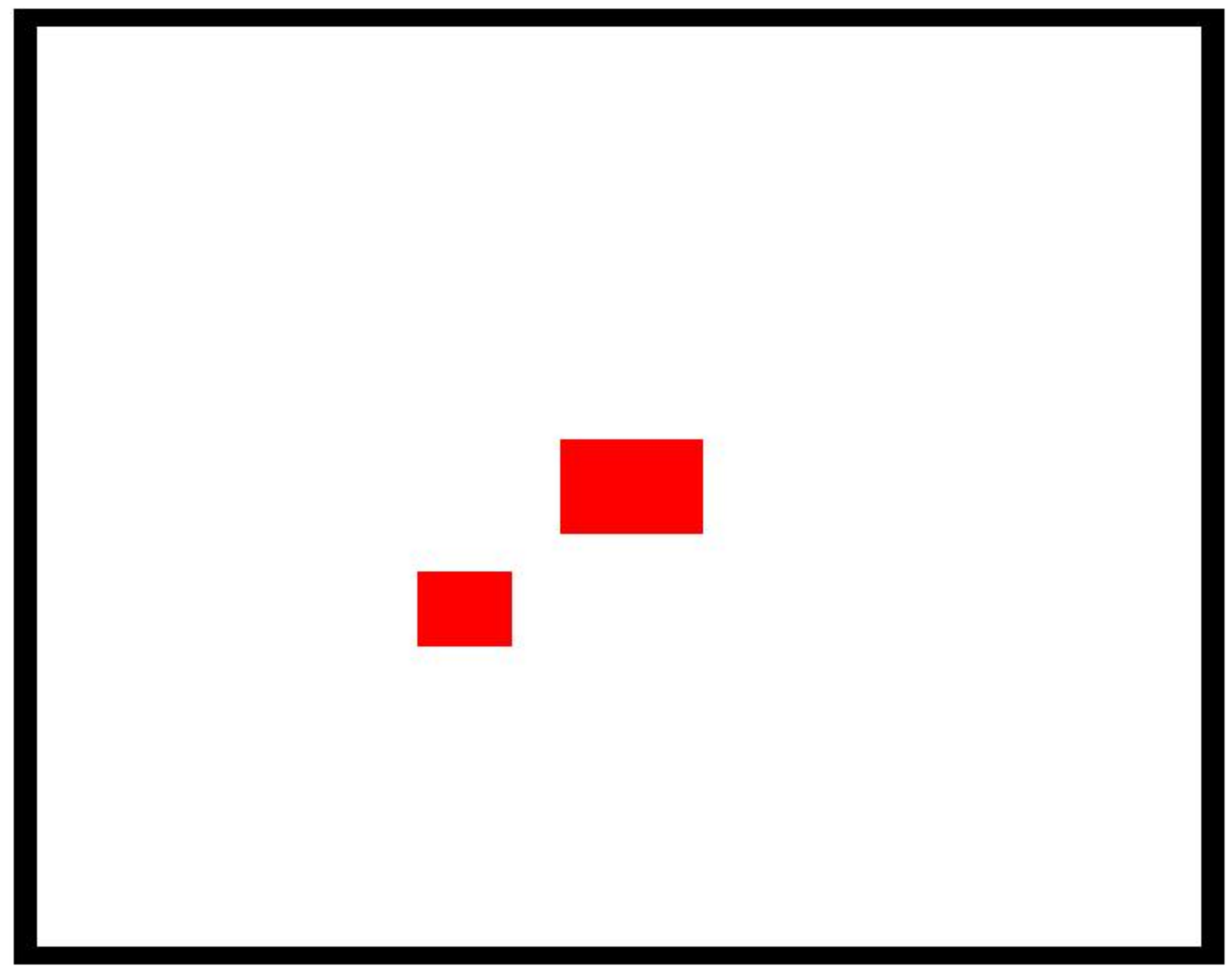} & %
\includegraphics[height=1.0 in, width =1.0 in]{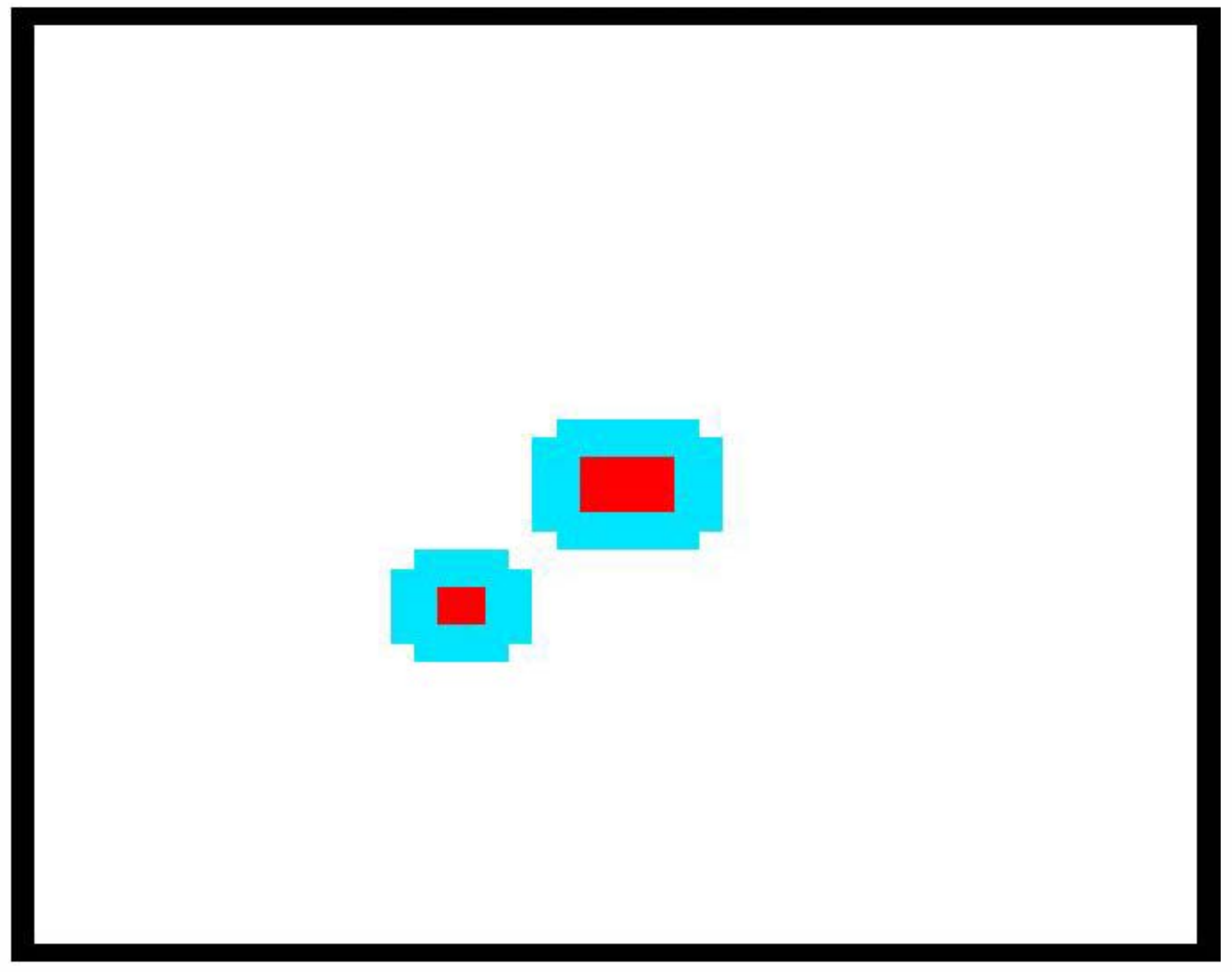}\\
(a) $t=0$ & (b) $t=0.5$ \\
\includegraphics[height=1.0 in, width =1.0 in]{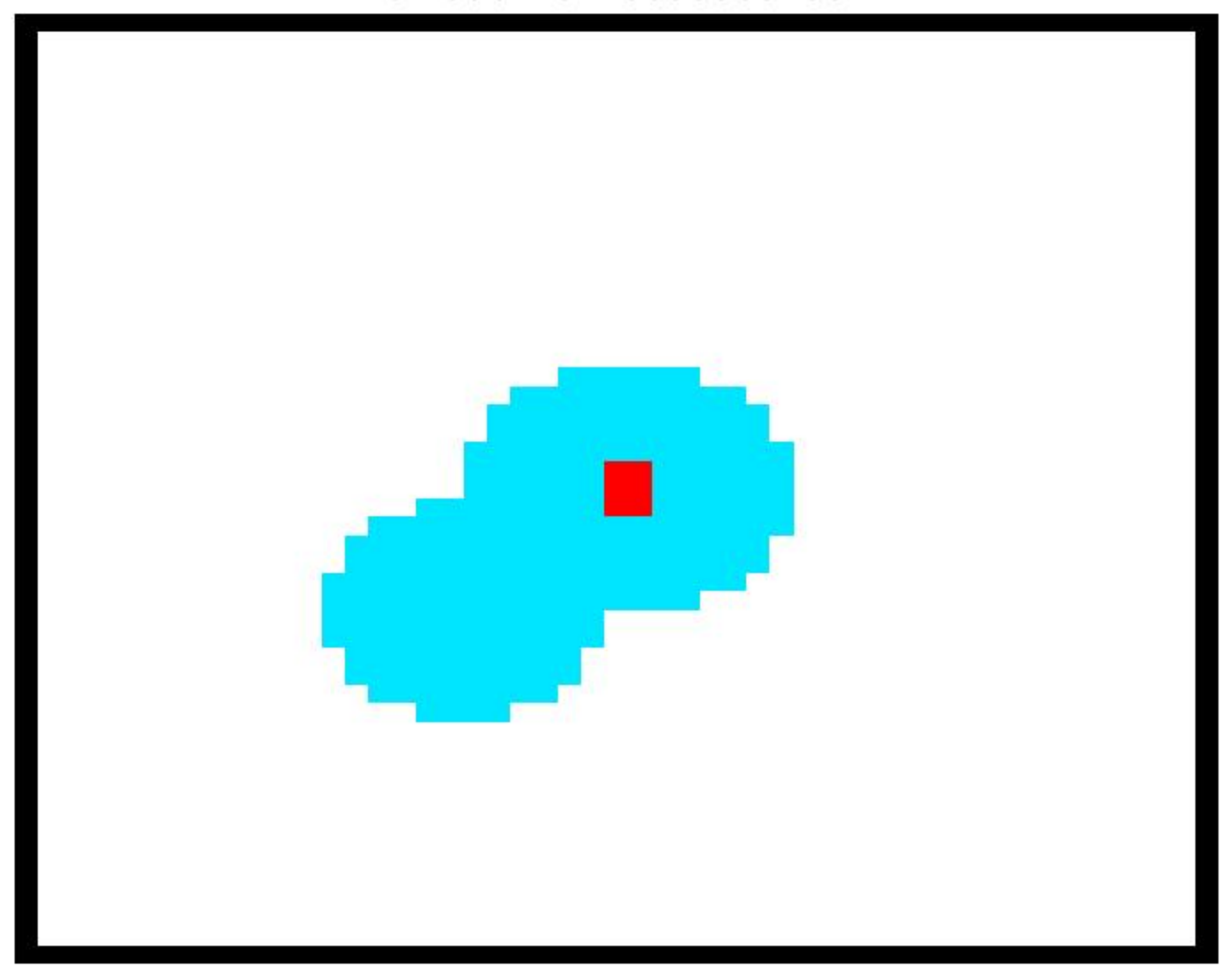} & %
\includegraphics[height=1.0 in, width =1.0 in]{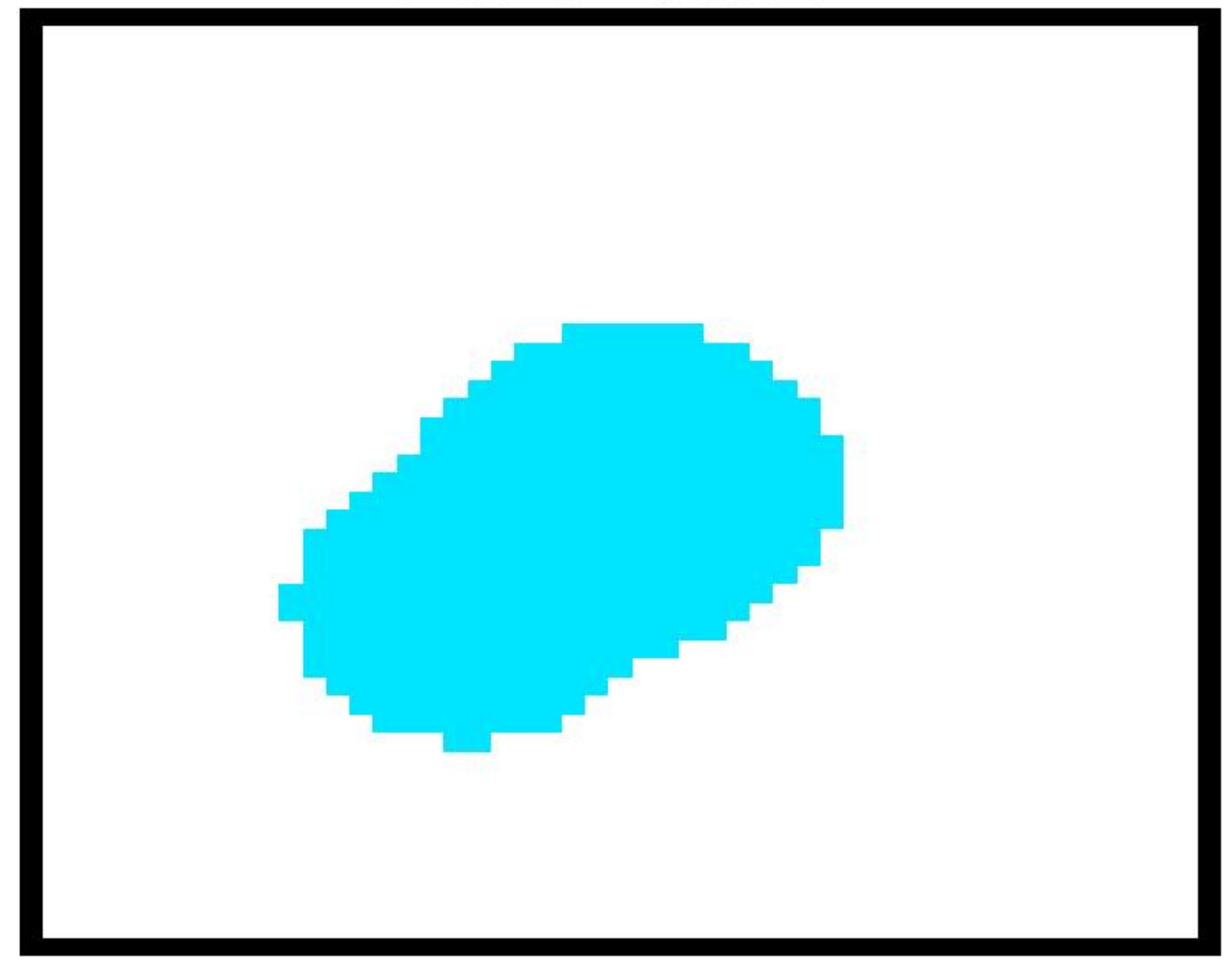} \\ 
(c) $t=5$ & (d) $t=12.5$
\end{tabular}%
\end{center}
\caption{Early evolution of a pair of excited rectangles. (a) is the 
initial condition, consisting of two spots of excitation, neither of which supports
a wave by itself. Making use of Figure \ref{phaseplane1} and the accompanying color code, 
we see in 
(b), (c), and (d) that all points in either spot follow PS curves
similar to that in Figure \ref{phaseplane2}(b), at least initially. It is also
seen that at none of these
points has $u$ become negative, as happens in Figure \ref{phaseplane2}(b) eventually. 
Further explanation is 
in the text. Here and in the next three figures, $b=0.1765$.}
\label{2spots:initial}
\end{figure}

\begin{figure}[htbp]
\begin{center}
\begin{tabular}{cc}
\includegraphics[height=1.0 in, width =1.0 in]{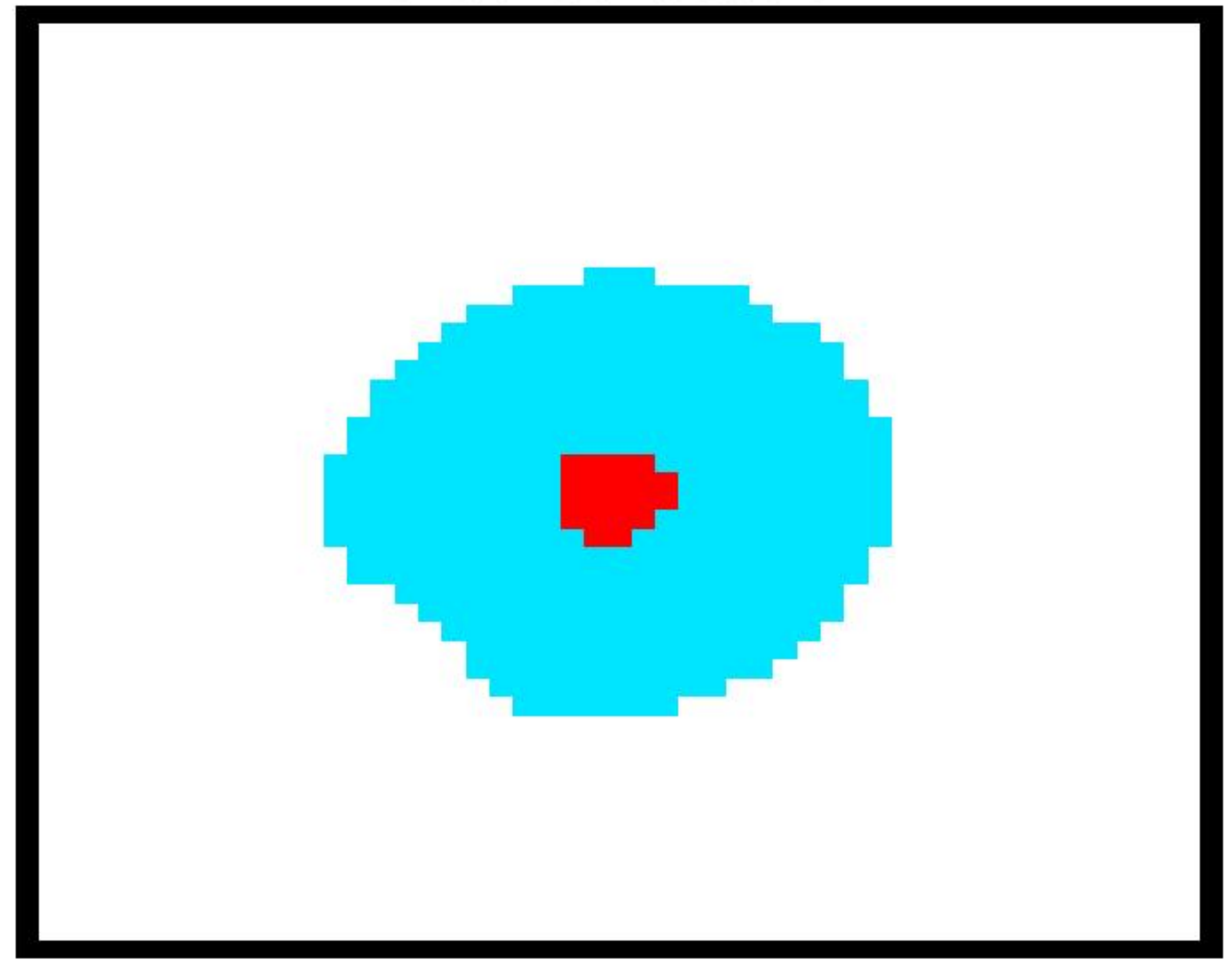} & %
\includegraphics[height=1.0 in, width =1.0 in]{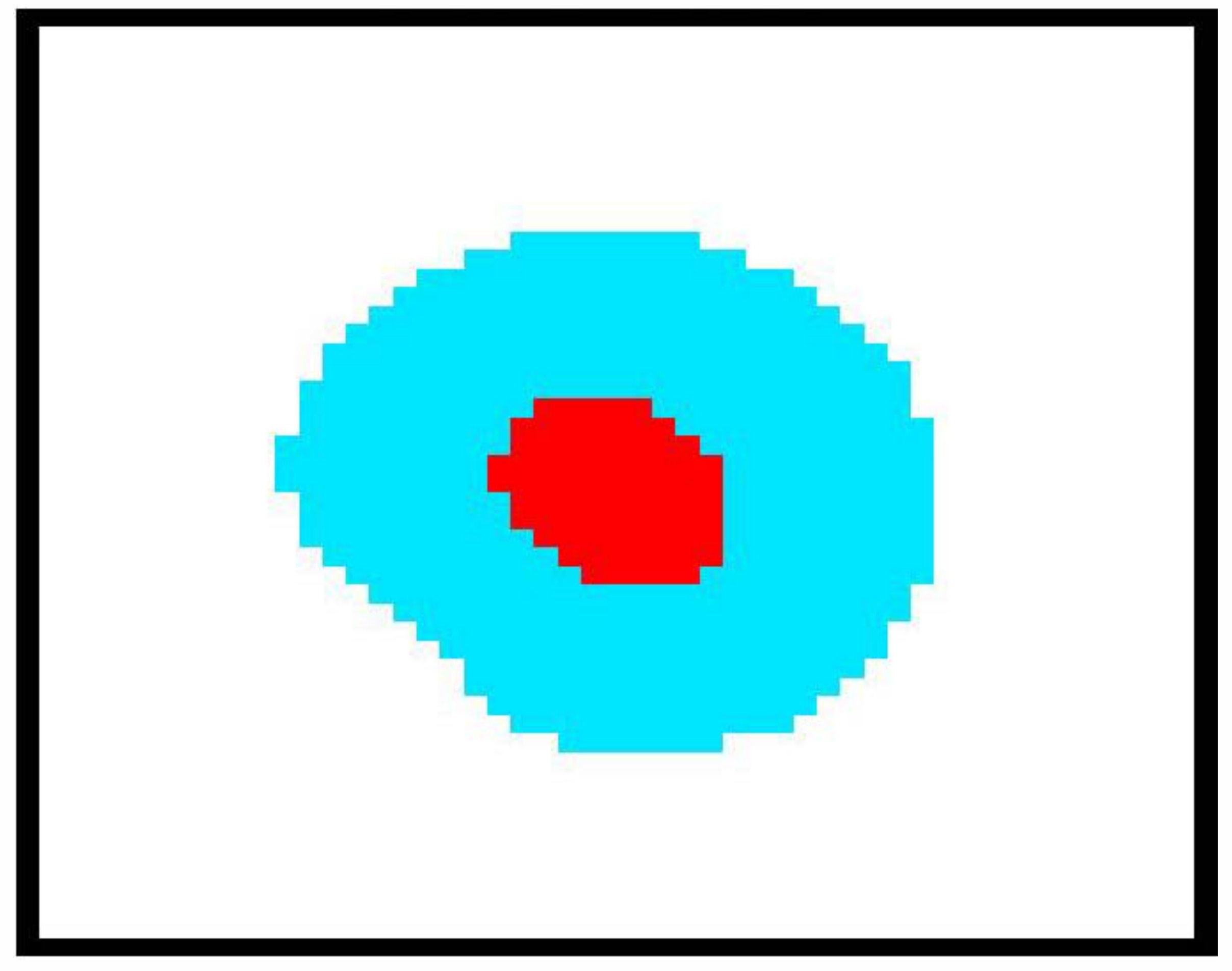} \\
(a) $t=28$ & (b) $t=78$ \\
\includegraphics[height=1.0 in, width =1.0 in]{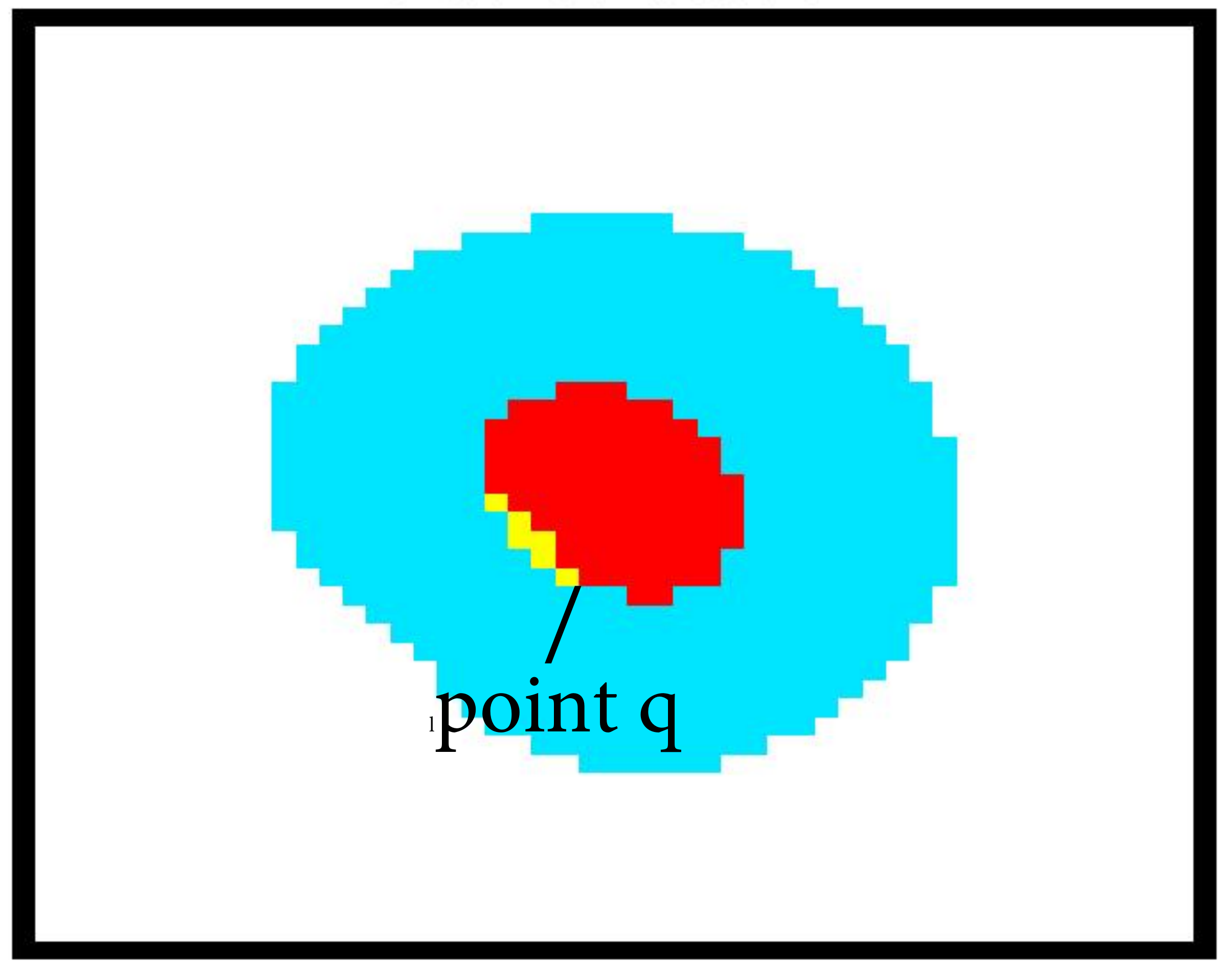} & %
\includegraphics[height=1.0 in, width =1.0 in]{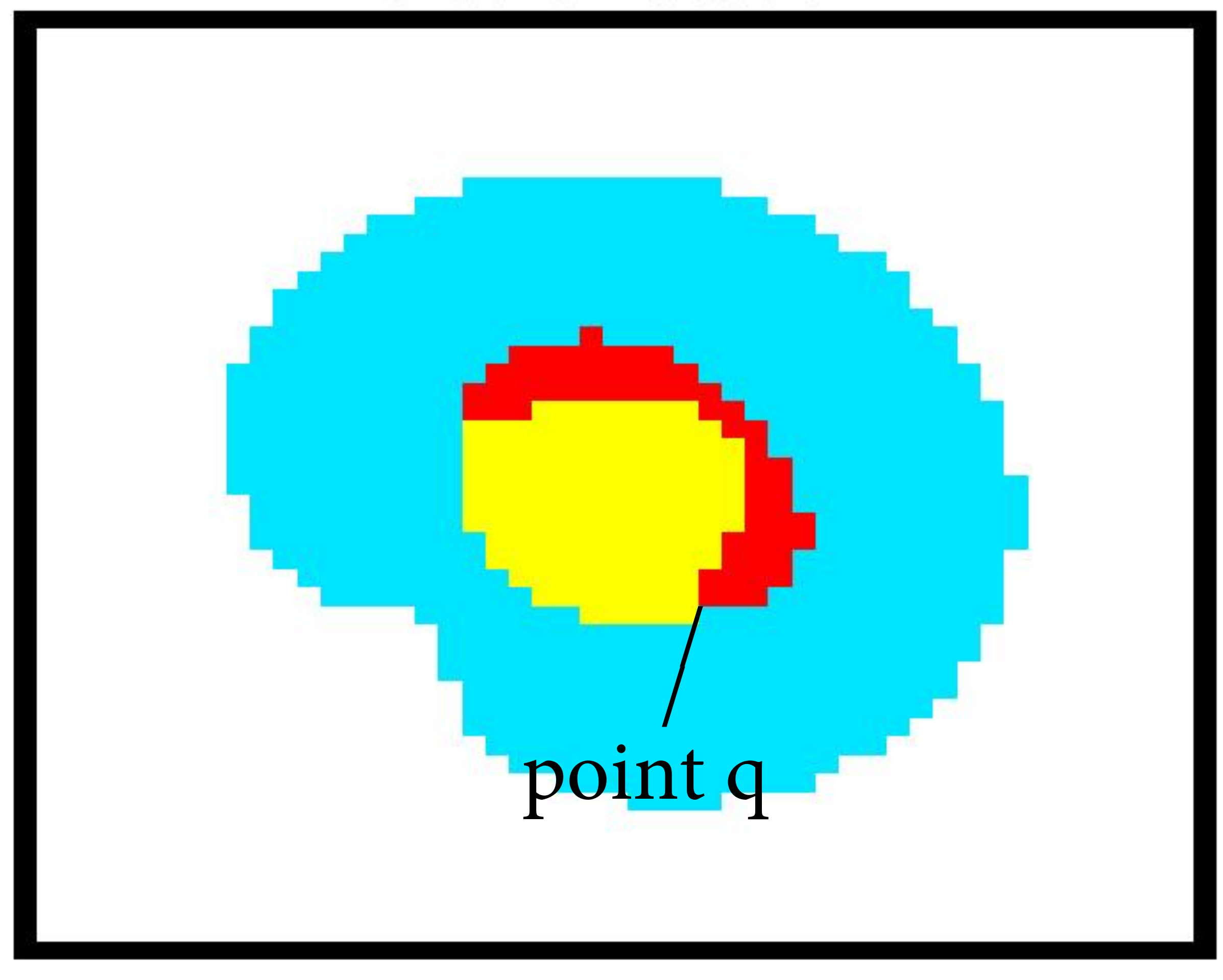} \\ 
(c) $t=88$ & (d) $t=120.5$
\end{tabular}%
\end{center}
\caption{Excitation in the central region. In (a), excitation has reappeared in
the center of the ring. In (a), (b), and (c) note the flattened 
boundary of the light blue region. In (c) the first yellow appears. 
In (d) a C-shaped region of excitation
has emerged, and with it, Zykov's point $q$, defined in the text. 
But still, no refractory region 
has appeared.}
\label{2spots:excited}
\end{figure}

\begin{figure}[htbp]
\begin{tabular}{ccc}
\includegraphics[height=1.0 in, width =1.0 in]{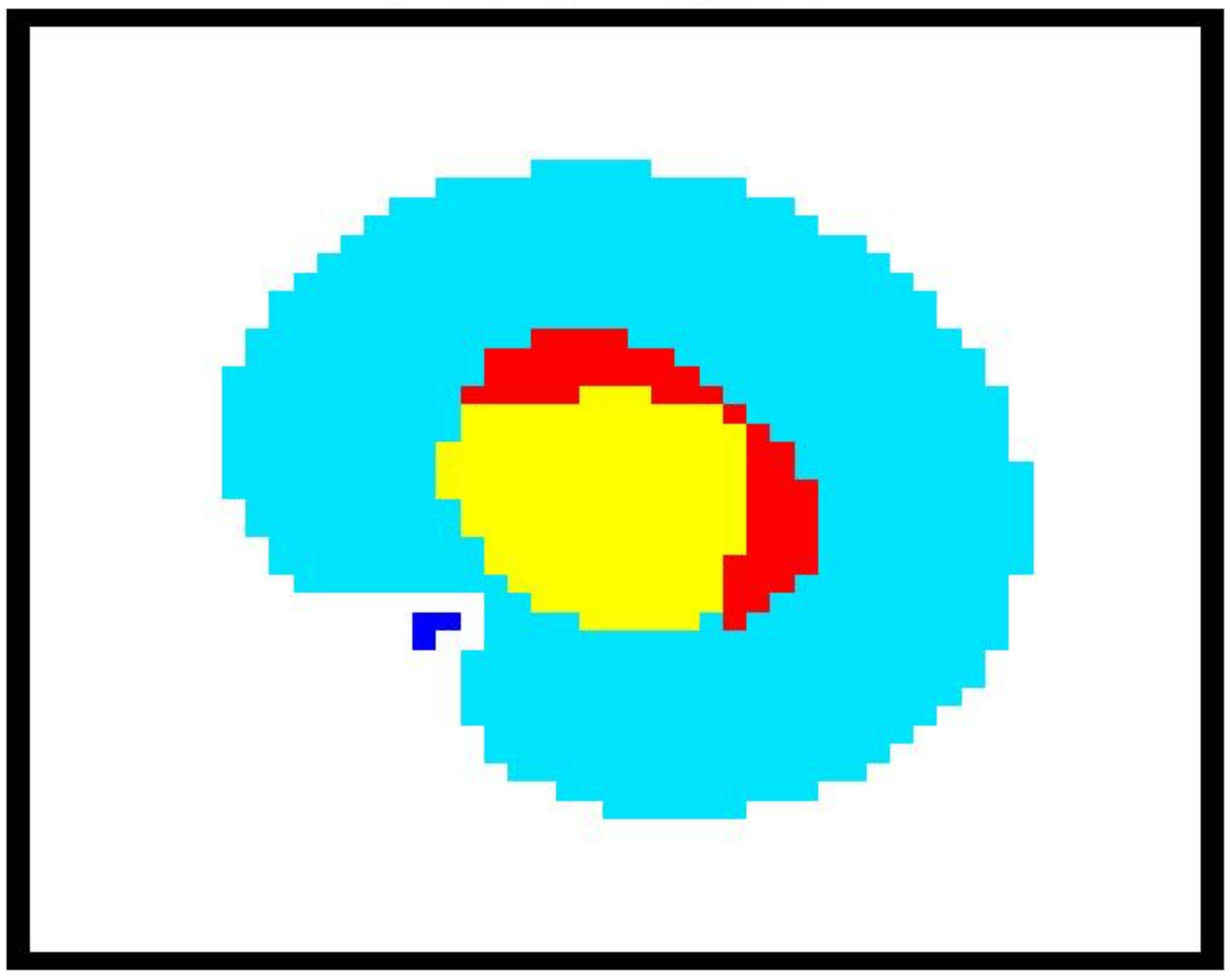} & %
\includegraphics[height=1.0 in, width =1.0 in]{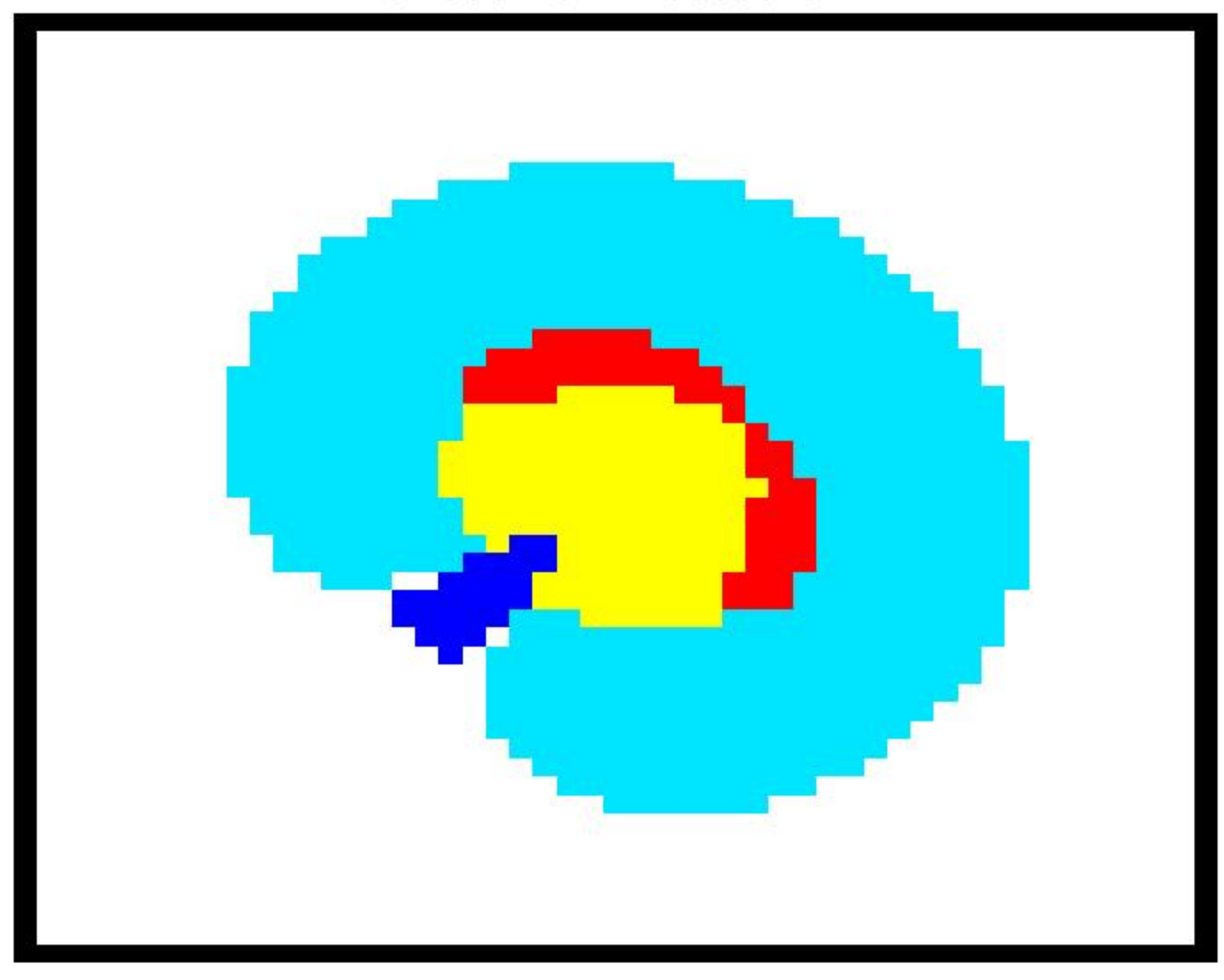} & %
\includegraphics[height=1.0 in, width =1.0 in]{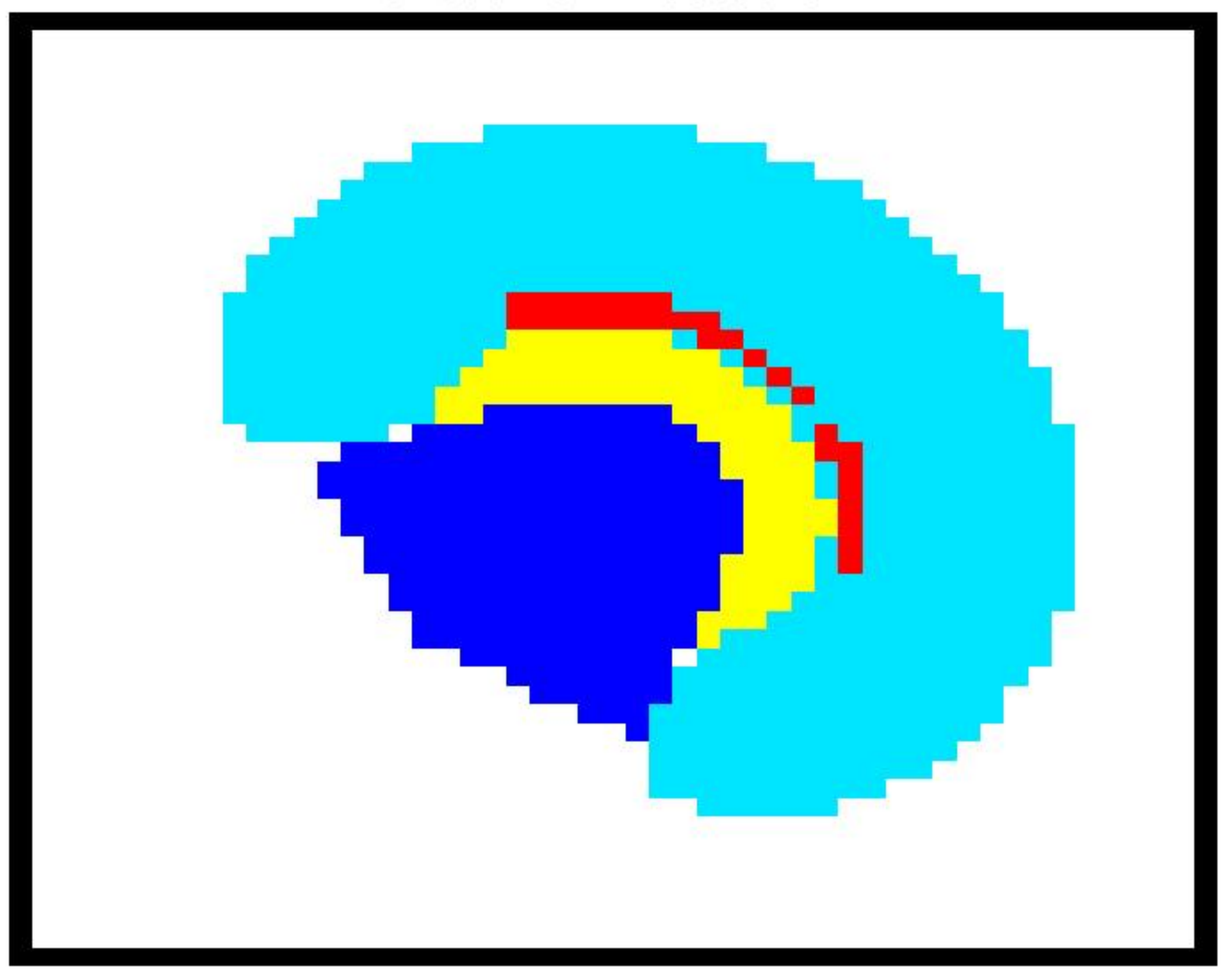} \\ 
(a) $t=123$ & (b) $t=125$ & (c) $t=147.5$ \\[8pt] 
&  & 
\end{tabular}%
\caption{In (a) we see the first sign of a refractory region. In (b) it 
has broken the blue ring and spread into the interior and in (c) 
the dynamics which carry trajectories from the wave back to the refractory 
region have led to a greatly expanded dark blue area. A moving C-shaped wave (the red,
light blue, and yellow regions) is now 
established. It is now appropriate to identify light blue as the wave front and yellow as the wave back, while
recognizing that these meet at the tips of the wave.	}
\label{2spots:wave}
\end{figure}

This is the example in Figure \ref{twospotsinitial}. Figures 
\ref{2spots:initial}-\ref{2spots:wave} show steps intermediate
to those in Figure \ref{twospotsinitial} in the evolution of
the two spots from initial configuration (\ref{2spots:initial}(a))
to a propagating $C$-shaped wave (\ref{2spots:wave}(c)).

Figure \ref{2spots:initial}(a) shows the initial condition. Figure \ref%
{2spots:initial}(b) shows the field after one numerical time step, when the
corresponding exact solution is smooth. Once a wave develops there will be
both light blue and yellow regions, with light blue the front and yellow the
back, but at this stage no yellow appears. There is a
lot of light blue, caused by diffusion from the rest regions acting on the
excited regions to lower the variable $u$ before $v$ has increased
significantly. Figure \ref{2spots:initial}(c) and (d) show the whole initial
pair of spots pulled into light blue. There is still no sign of either
yellow or a refractory region.

\begin{figure}[htbp]
\begin{center}
\begin{tabular}{cc}
\includegraphics[height=1.6 in, width =1.6 in]{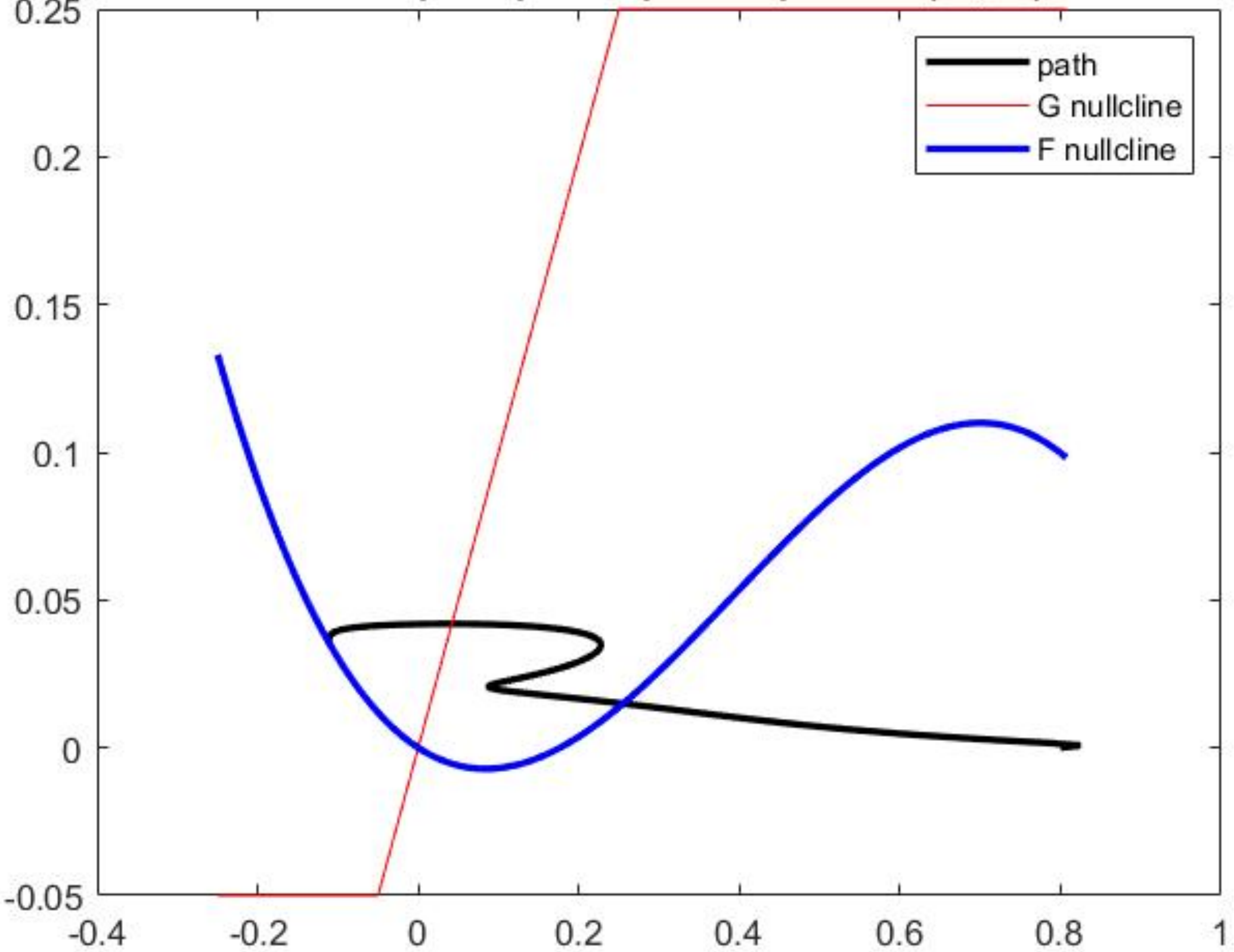} & %
\includegraphics[height=1.6 in, width =1.6 in]{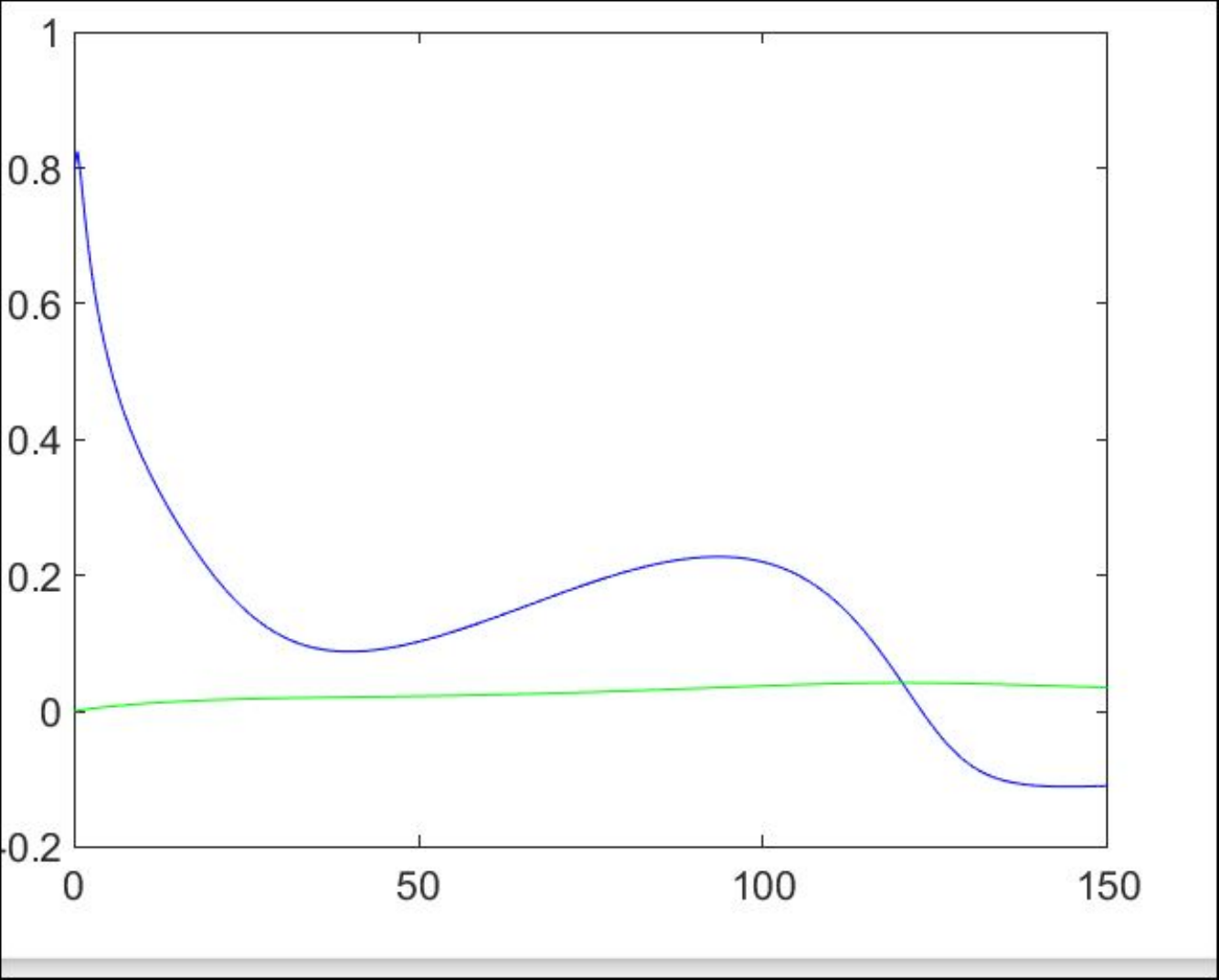} \\
(a)   & (b)   \\
&
\end{tabular}%
\end{center}
\caption{ (a) The curve $ \rightarrow  (u(x,y,t),v(x,y,t)) $ for a point $(x,y)$ in the
smaller spot in Figure \ref{twospotsinitial}(a). (Compare with Figure \ref{phaseplane2}(b).)  (b) The graph of $u(x,y,t)$ as a function 
of $t$ for the same point $(x,y)$ as in (a). In the ``switchback'' region of (a), corresponding to the up and down motion of 
$u(t)$ in (b), diffusion and reaction are closely balanced, and this oscillation delays 
the arrival of $(u,v)$ in the refractory region. This may contribute to the flattening
of the light blue region in Figure \ref{2spots:initial} (a), (b), and (c).}
\label{2spots:ps}
\end{figure}

Figure \ref{2spots:excited}(a) shows reestablishment of an excited region.
Here $v$ is still low, so that the dynamics
take over and return the middle of the original larger region to the excited
state. Also there is a slight flattening of the blue region on the left and
this becomes more pronounced in \ref{2spots:excited}(b).  This flattening
is likely due to the now unseen smaller rectangle, and in Figure \ref{2spots:ps} below we 
propose a possible mechanism, but the exact
process needs more study. Note that there are not yet any refractory cells.

In \ref{2spots:excited}(c) and (d), $v$ is large enough to give a small
yellow region, caused by the dynamics since the points involved have made
the standard transition \ red $\rightarrow $ yellow. This yellow region
makes the excited region concave nearby, giving the first hint of a
$C$ shape. By (d) the excited region has a clear $C$ shape, yet still, no
refractory areas are evident.

A common description of what causes repetitive patterns, perhaps originating
in \cite{Wiener}, is that the front and the back of a wave meet.
This is very apparent in Figure \ref{2spots:excited}(d).
The light blue and yellow regions meet along a curve. Moreover, there
is now a point at the end of each red tip where red, yellow, and light blue
come together. Such a point has been emphasized by Zykov \cite{Zykov1} as
crucial for movement of a fixed rotating spiral. He calls a point of this
kind \textquotedblleft point $q$\textquotedblright.

Figure \ref{2spots:wave} shows the establishment of a moving $C$-shaped
wave. A refractory region (dark blue) has appeared and broken through the
light blue ring. This refractory region has not yet been either red or yellow,
and so has not followed the standard order of colors. Hence it must have been
caused by diffusion, at least up to the point that $\left(
u,v\right) $ cross the nullcline branch $\gamma$. 

Given a large enough region, considerably larger than that shown in Figures \ref{2spots:initial} - 
 \ref{2spots:wave},  the $C$ shaped wave in this example grows and appears as if it will become a
closed curve, but the tips of the $C$ curl around and merge, with
flared out ``lips'' which then break off to form a tiny region that grows into
its own $C$ shaped wave, while the outer closed wave grows outward and leaves
the computational region.  This repetitive process is described in more
detail in the next example and shown in Figure \ref{CurledLips} and an associated movie.

We should mention, however, that formation of a C-shaped wave does not guarantee that the process will become repetitive.  The appearance of point $q$ makes lasting patterns possible, but nothing is
guaranteed at this stage. Point $q$ could disappear; we saw this frequently, and in this example we 
corrected it by moving $b$ further into the weakly excitable region.  The result is shown in a movie after the next example.

We are far from understanding exactly what features are important in this example and others to follow. The timing
of the refractory region seems to be important. In the following figure we show a PS curve
in the area where such a region first arises and also plot $u$ vs time on this curve.

\subsubsection{Views using unfilled contours}

\label{contours}

\begin{figure}[H]
\begin{center}
\begin{tabular}{cc}
\includegraphics[height=1. in, width =1. in]{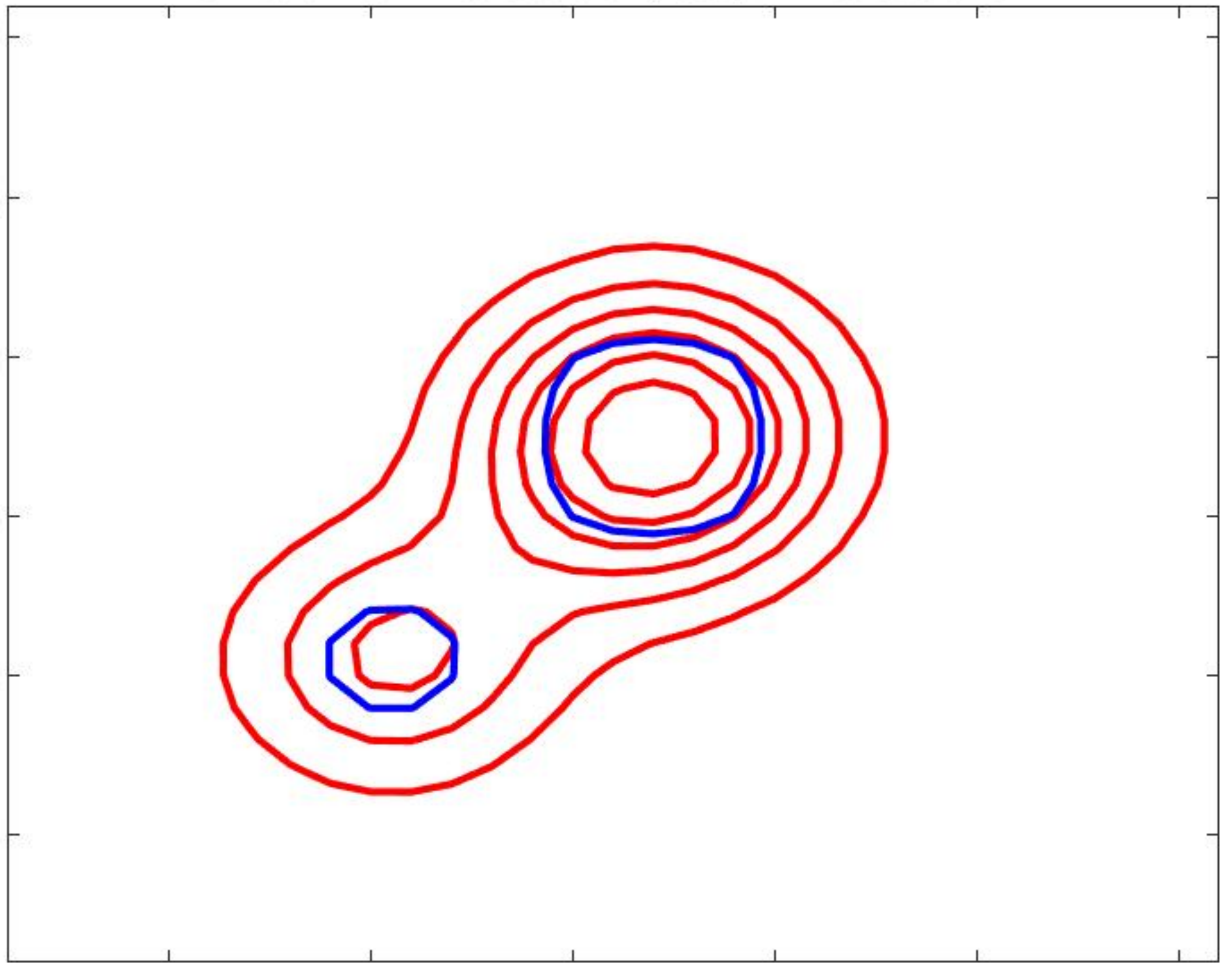} & %
\includegraphics[height=1. in, width =1. in]{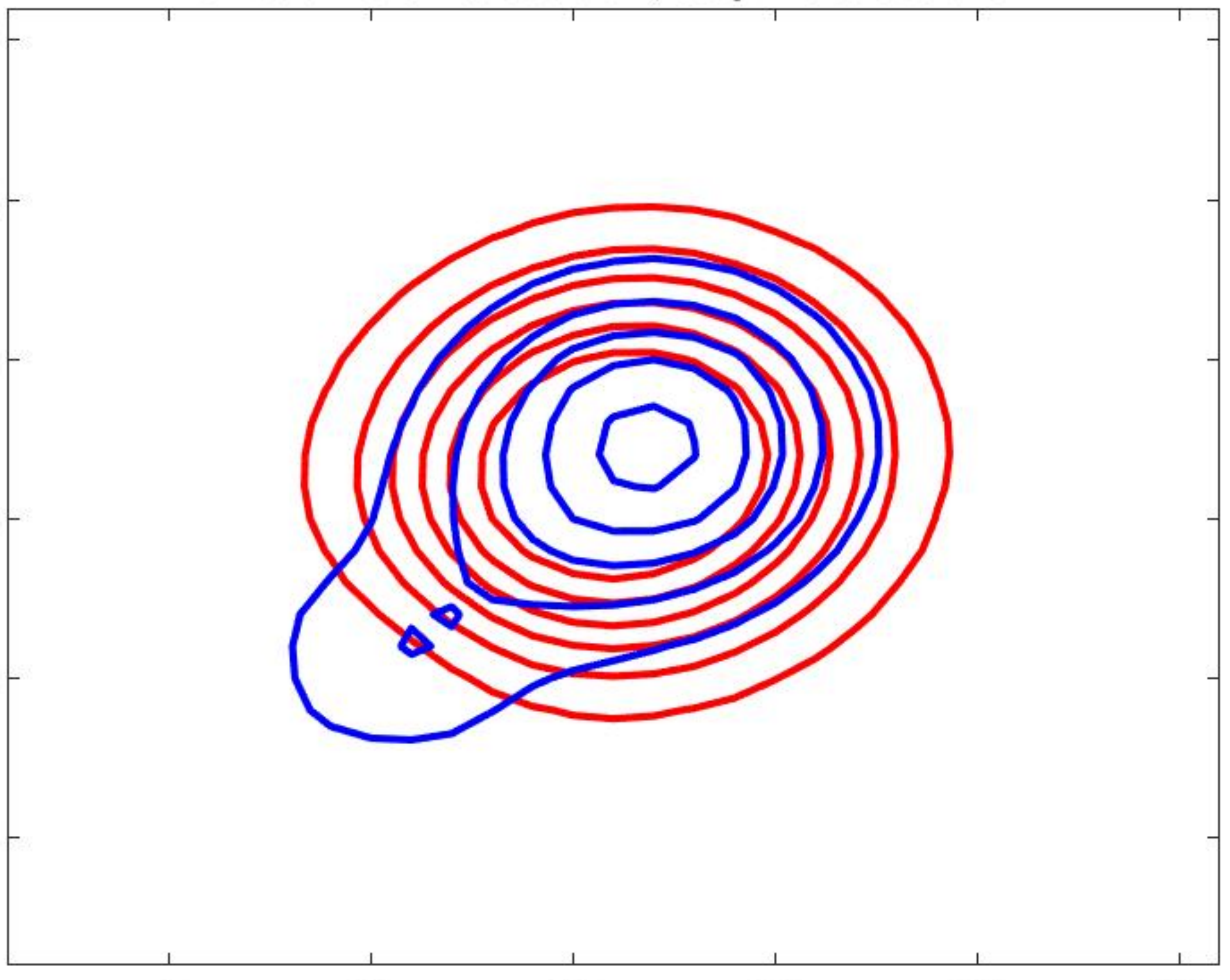}\\
(a) $t=12.5$ & (b) $t=38$ \\
\includegraphics[height=1. in, width =1. in]{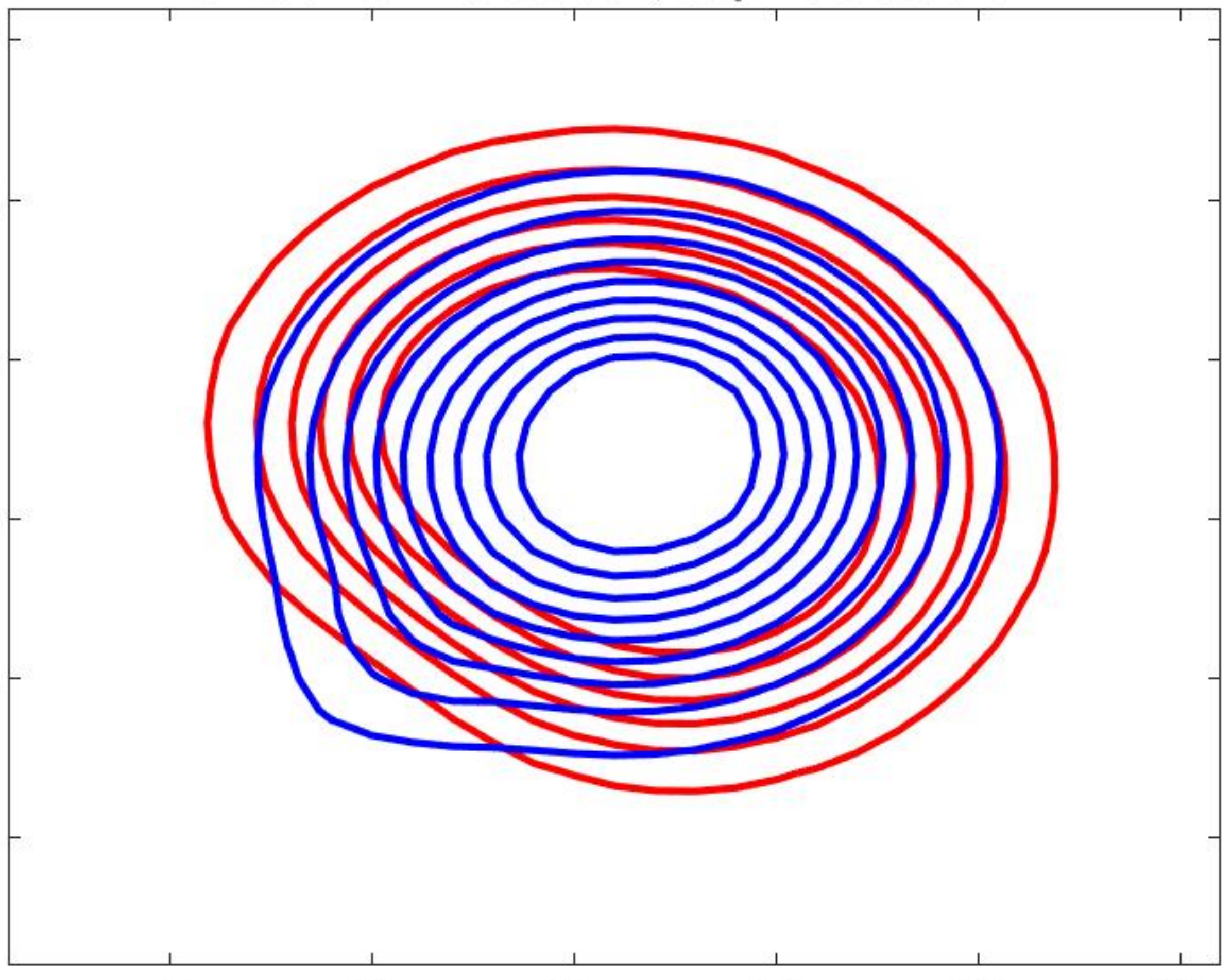} & %
\includegraphics[height=1. in, width =1. in]{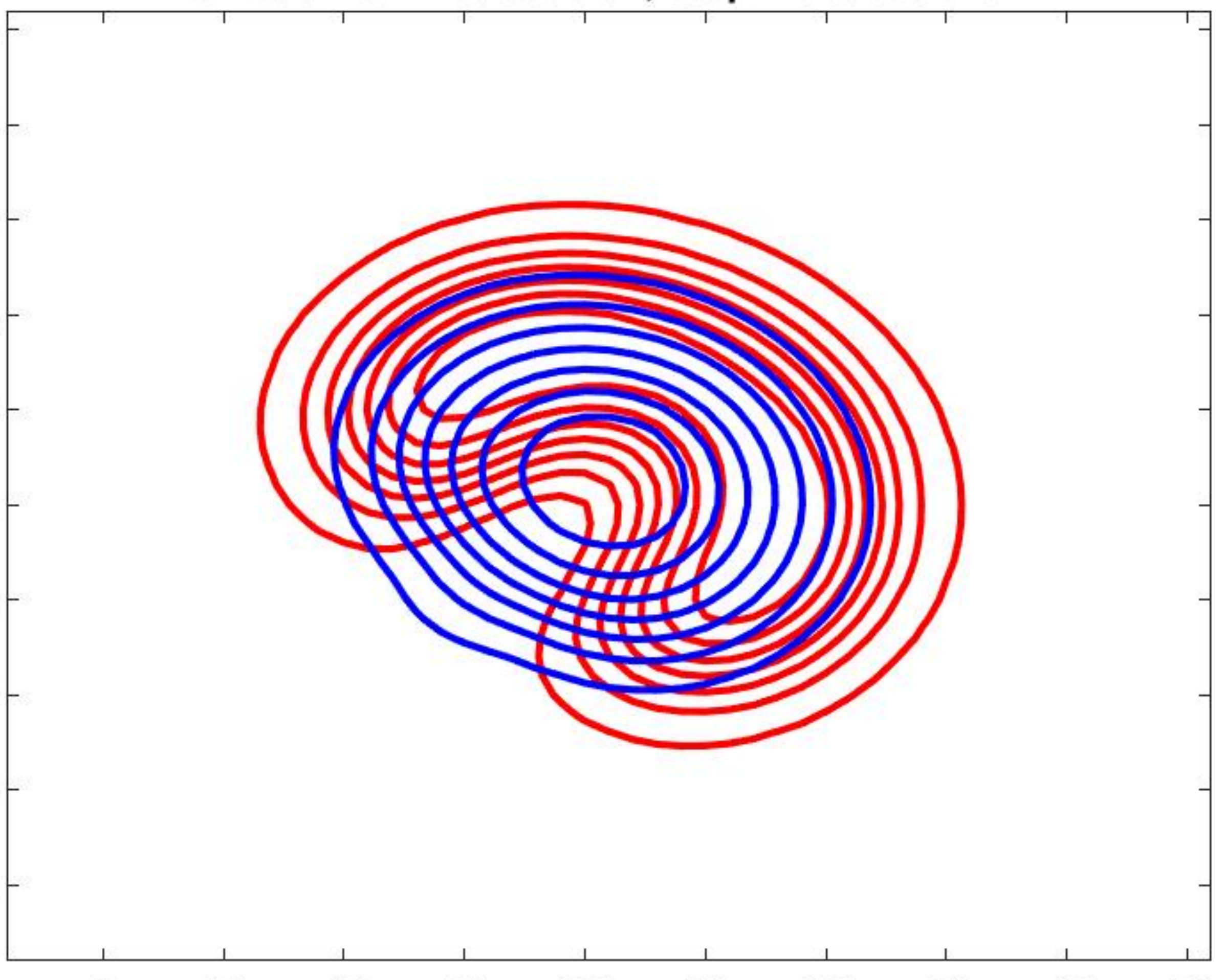} \\ 
(c) $t=78$ & (d) $t=125.5$ \\
&
\end{tabular}%
\caption{Contour versions of Figures \protect\ref{2spots:excited}(d) and 
\protect\ref{2spots:wave}(a),(b),(d); red: $u$ contours, blue: $v$ contours.
At the center of a nest of red contours, $u$ is high and at the center of a 
nest of blue contours, $v$ is high. In (d) most of the $u$ contours cross most of the 
$v$ contours, and this results in point $q$ appearing, at a point in the cross-hatched region. 
Compare with Figure \ref{2spots:wave}(b). See
https://pitt.box.com/s/ehu6dgm5citma3sniyrsb3twdmy9qsr3. 
}
\label{contour}
\end{center}
\end{figure}

More information can be gleaned by looking at contour plots of $u$ and $v$
together. Figure \ref{contour} repeats parts of Figures \ref{2spots:initial}
and \ref{2spots:excited} in contour form.

Up to around \thinspace $t=12.5$ the levels curves of $u$ and $v$ track each
other, but later, these two sets of curves intersect in this region. \
Specifically, most of the trajectories of $u$ cross most of the trajectories
of $v.$ This is seen clearly in two patches in Figure \ref{contour}(d),
where six levels curves of $u$ cross six level curves of $v.$ In the center
of such a patch the $u$ level curves are particularly close together,
indicating rapid change of $u$ in a direction crossing these curves. \ The $(u,v)$ values
at points $(x,y)$ in the central part of the cross-hatched region are in an 
area  of the $(U,V)$ phase plane which is highly unstable for the odes in (\ref{2.3},\ref{2.4}). 
 Zykov's point $q,$ two copies of
which are born between Figures \ref{contour}(c) and (d), is in this region.

\section{A second model giving rise to repetitive motion}
\label{symmetric}

After considerable experimentation, we were able to generate a
class of symmetric initial
configurations that would result in apparently sustained oscillations.

\begin{figure}[]
\begin{tabular}{ccc}
  \shortstack{\includegraphics[width=0.70in]{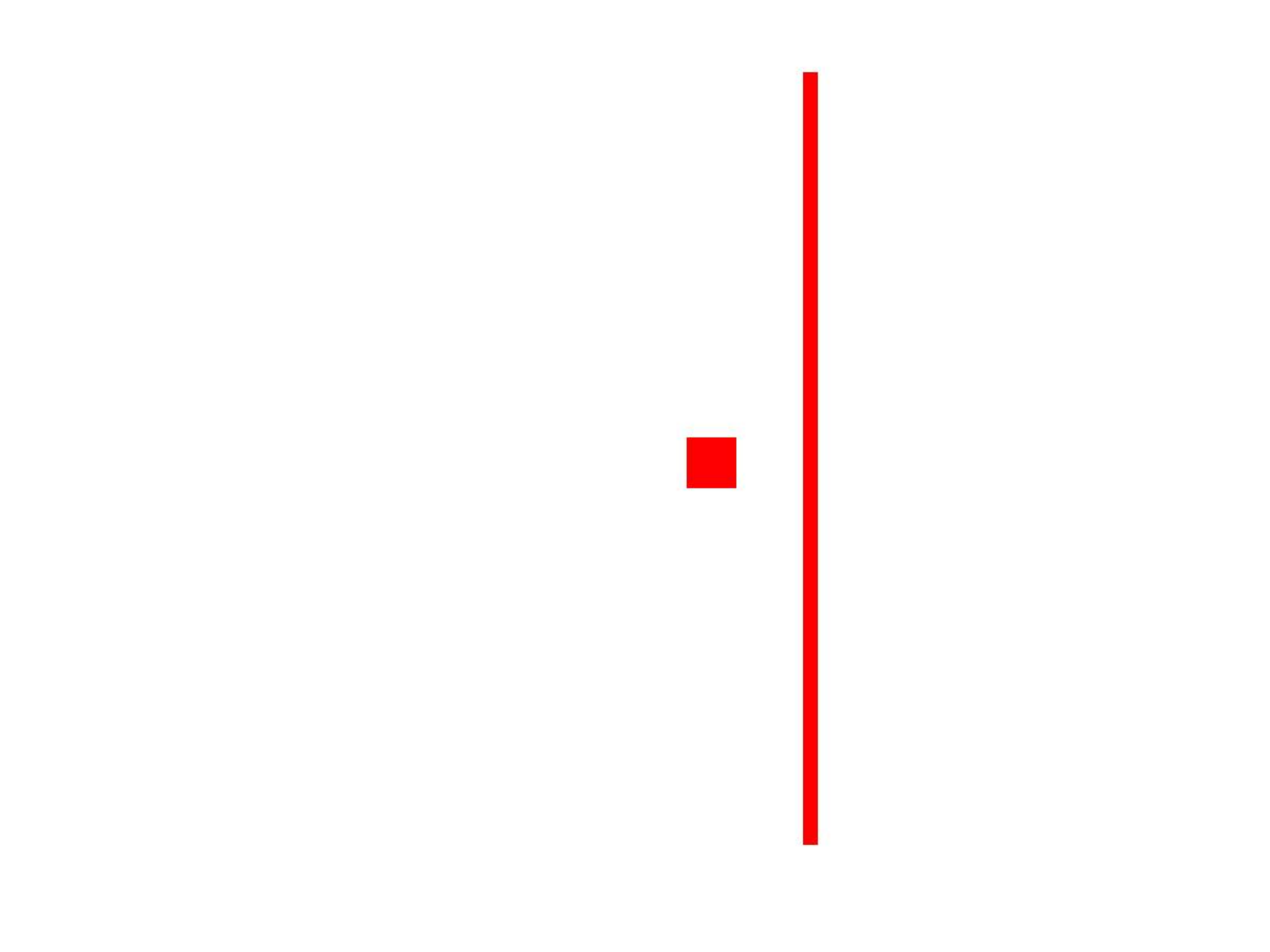}
    \\ \rule{0pt}{6pt}} &
\includegraphics[width=0.8in, height=0.8in]{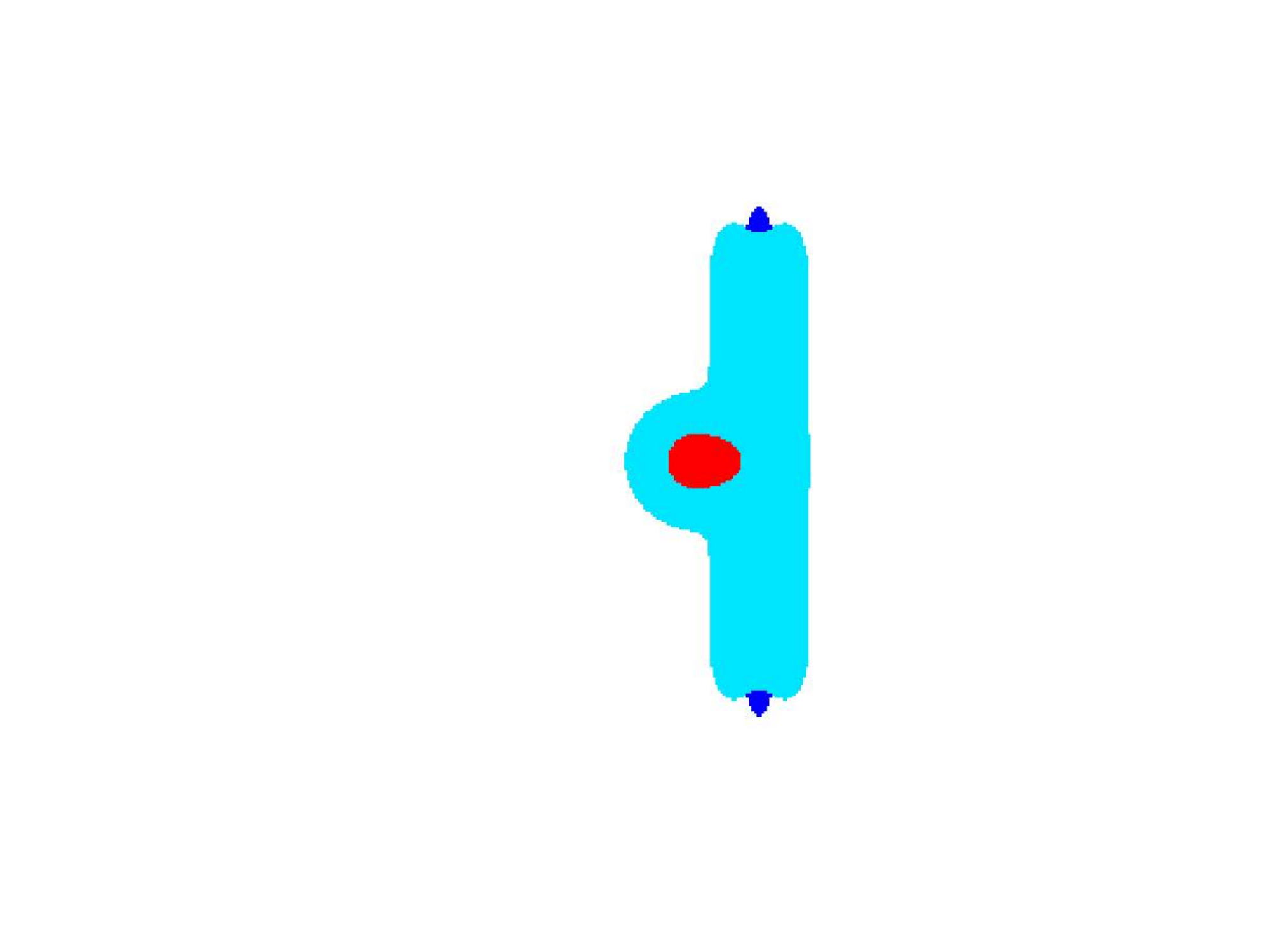} & %
\includegraphics[width=0.8in, height=0.8in	]{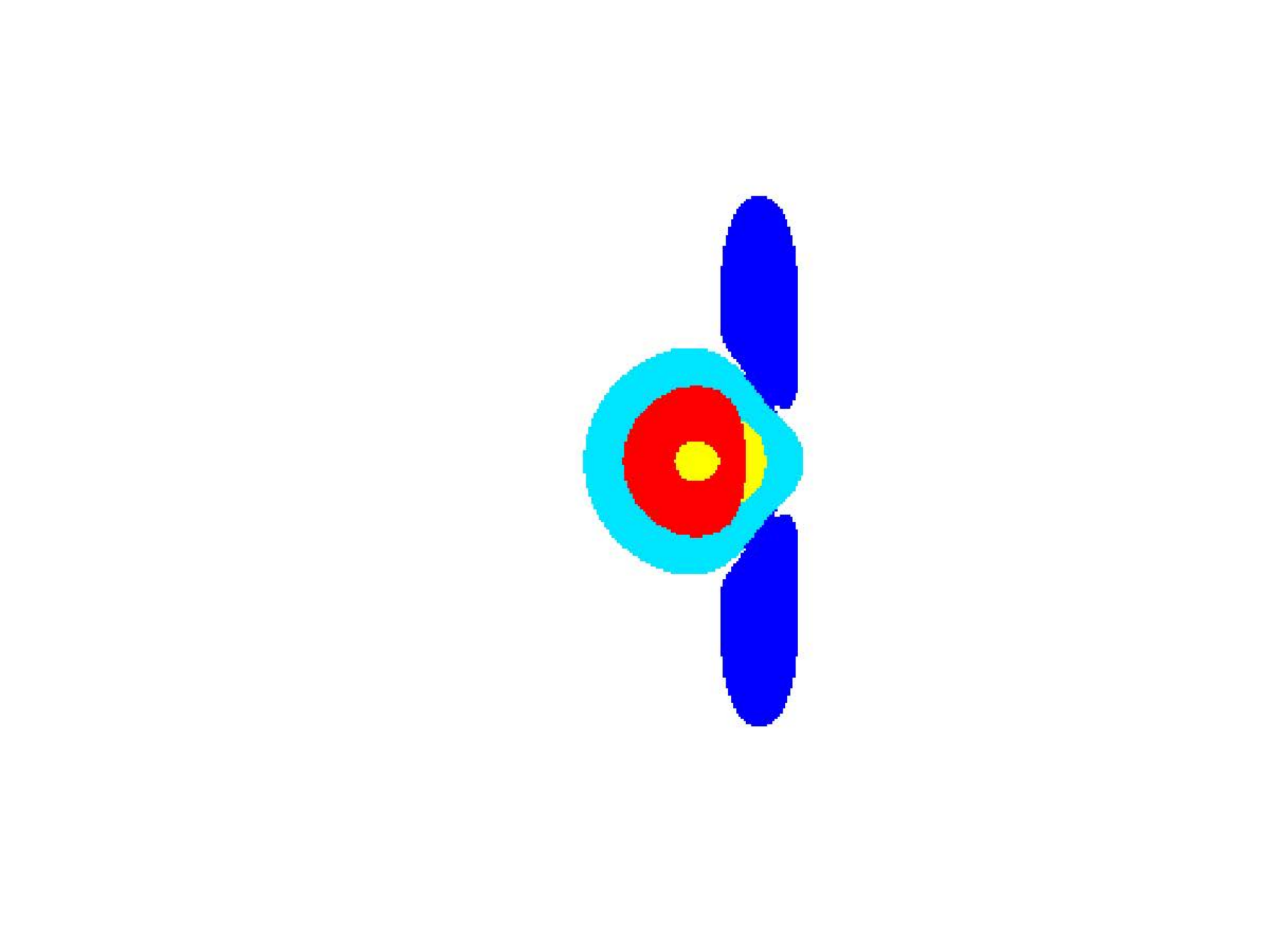} \\
(a) $t=0$ & (b) $t=40$ & (c) $t=102.5$ \\
\includegraphics[width=0.8in, height=0.8in]{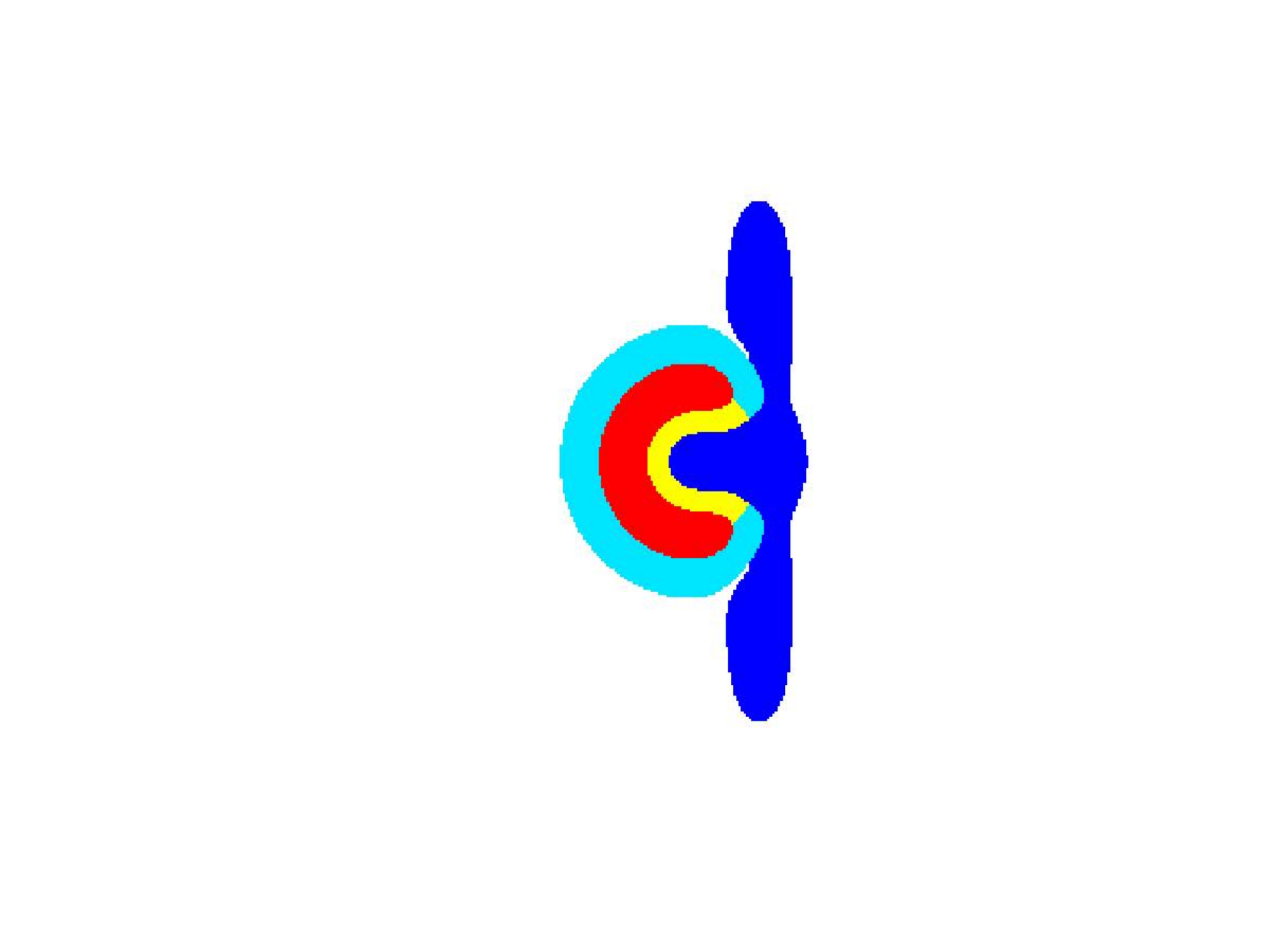} &
\includegraphics[width=0.8in, height=0.8in]{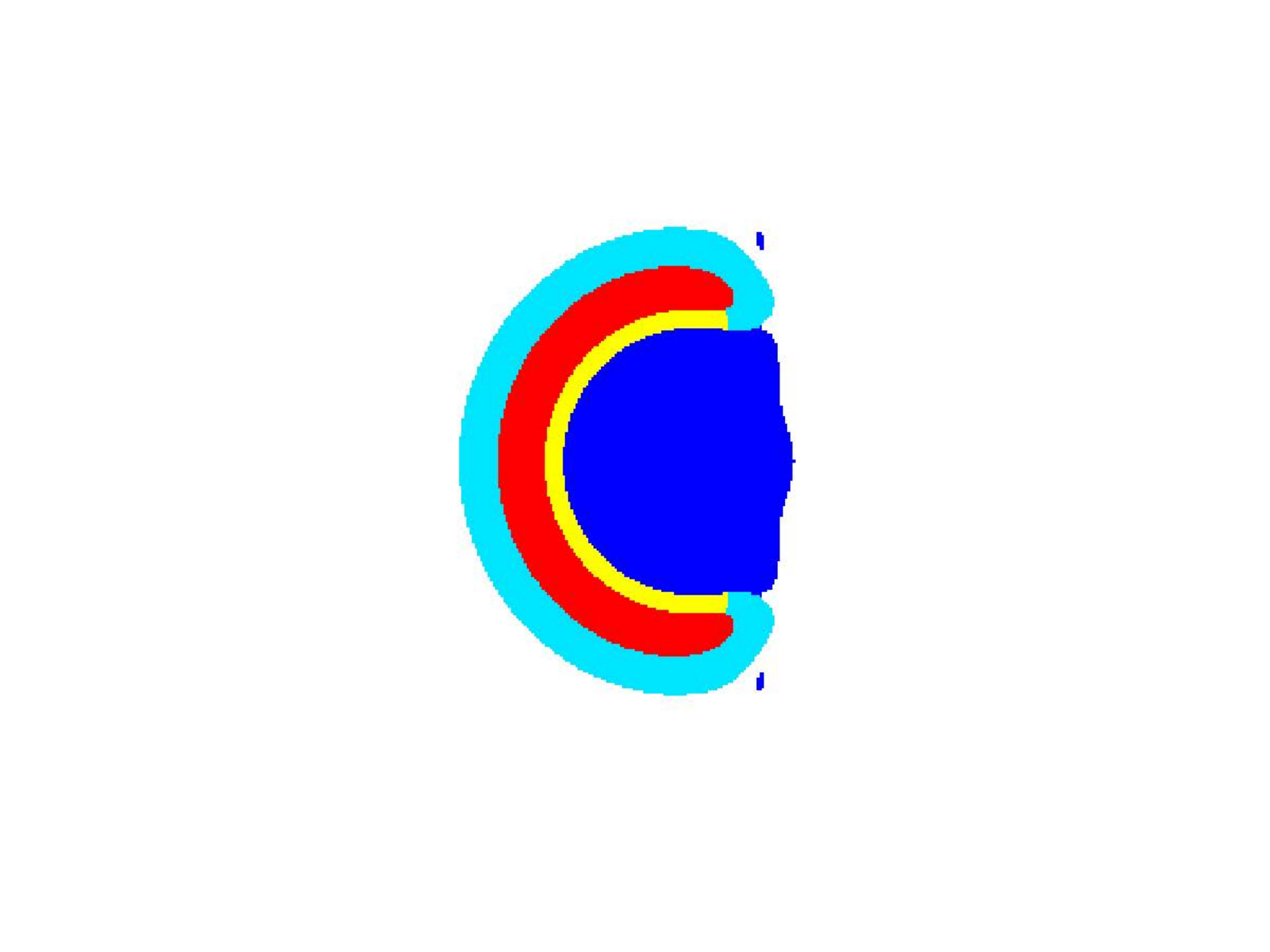} &
\includegraphics[width=1.0in, height=1.0in]{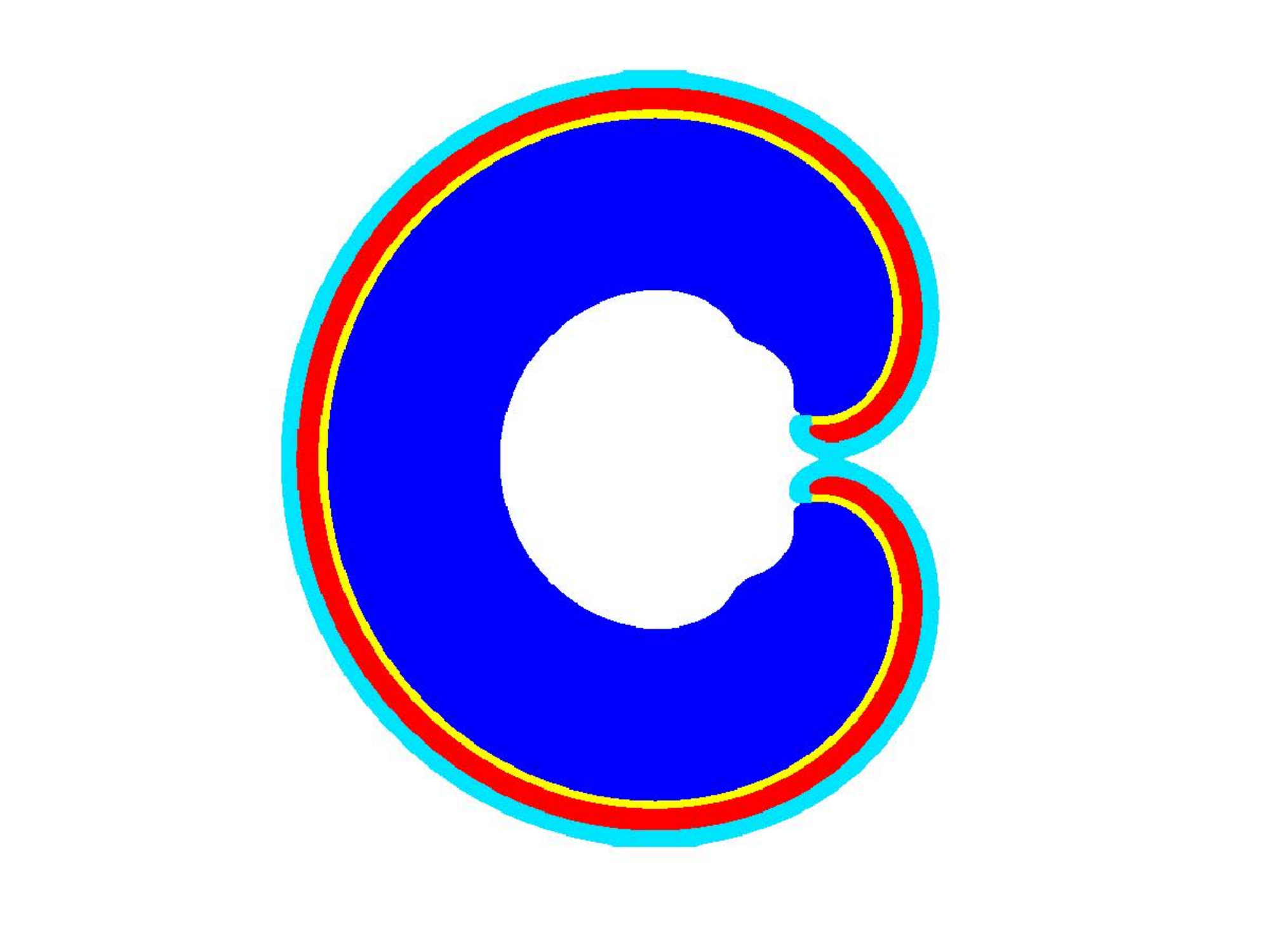}\\
(d) $t=132.5$ & (e) $t=250.0$ & (f) $t=987.5$
\end{tabular}%
\caption{(a) is the initial configuration; (b) and (c) show development
of the small spot $E$ into a target wave and the long thin barrier
$B$ becoming refractory;  (d) shows
the target wave broken into a C shaped wave; and (e) and
(f) show growth of the C shaped wave.  Here $b=0.17$. }
\label{sym:startup}
\end{figure}

The initial condition can be described as a
small square $E$ just large enough to generate an outgoing circular wave
placed to the left of a long thin \textquotedblleft
barrier\textquotedblright\ rectangle $B$. The barrier $B$ is subcritical.
Diffusion reduces its excitement faster than the dynamical terms
can raise it. This region
becomes refractory and serves to block expansion of the circular wave
generated from $E$. The circular wave is partially extinguished as it
expands to the space occupied by $B$ and is broken into the familiar
expanding $C$ shaped wave, as in Figure \ref{sym:startup}(e) and
\ref{sym:startup}(f).

\begin{figure}
\begin{center}
\begin{tabular}{cc}
\includegraphics[height=1.0in]{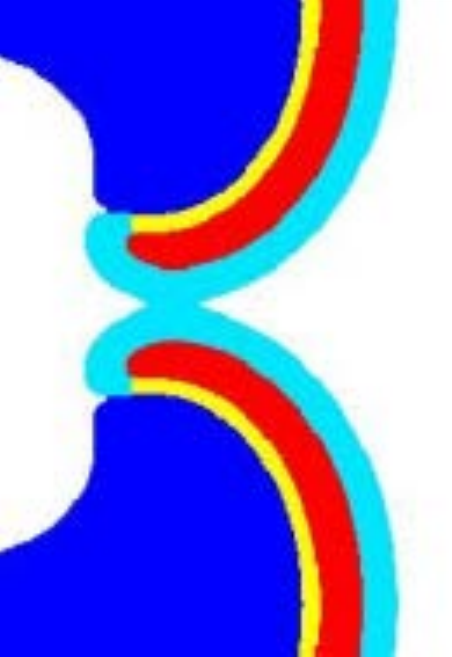} &
\includegraphics[height=1.0in]{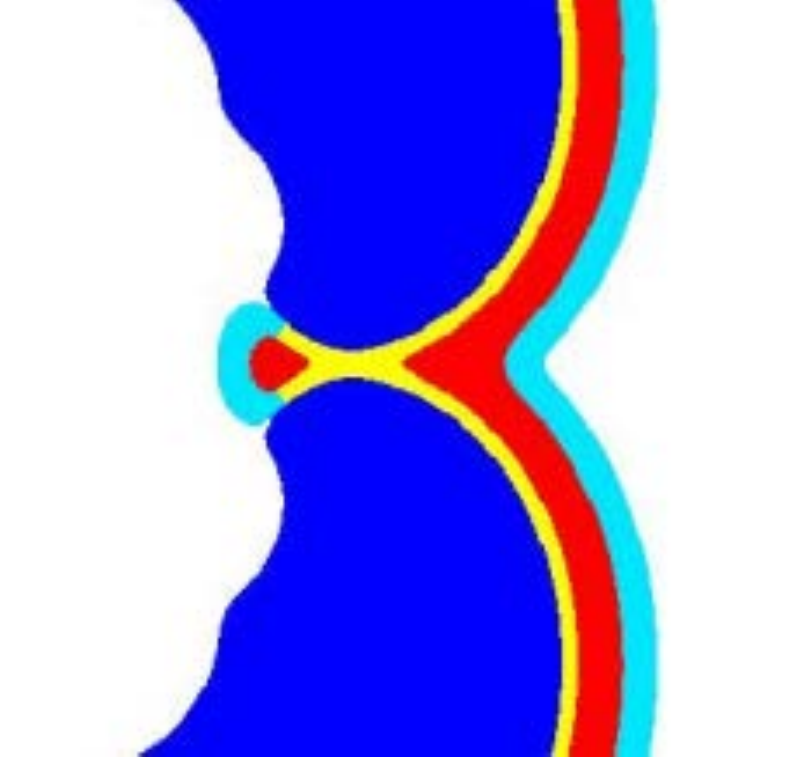}\\
(a) $t=987.5$ & (b) $t=1087.5$
\end{tabular}
\end{center}
\caption{Detail of (a) closure of outer ring showing curled
  ends and (b) detachment of those ends to eventually produce a
  new ring. \label{CurledLips}}
\end{figure}

If the barrier $B$ is short, the $C$ shaped wave reconnects and becomes an
outgoing circular wave. If $B$ is longer, the wave ends curl sufficiently to
leave a remnant as shown in Figure \ref{CurledLips}.  This remnant
may or may not remain sufficiently excited to continue to evolve, depending
on the length of $B$.

In detail, the initial configurations can be given as two rectangles, 
supercritical $E=[0.02,0.05]\times \lbrack -0.015,0.015]$ and 
subcritical $B_{\alpha}[0.095,0.1025]\times \lbrack -\alpha ,\alpha ]$, 
for $0<\alpha \leq 0.5$.
Both of these rectangles are used to set $u=0.8$ and $v=0$, values
representing purely excited regions. The supercritical $E$ rectangle is 
large enough that it, without $B_{\alpha }$ will
result in a circular outgoing target wave that allows all points to return to the
equilibrium state and remain there after the wave passes. For any $\alpha
\leq 0.5$, $B_{\alpha }$ is subcritical.
Values of $\alpha >0.5$ have not been checked.

Figure \ref{sym:startup} shows the solution at five early times, starting with 
the initial condition. Here $\alpha = 0.5$.
The small square region is an $11\times 11$ set of mesh
squares, each of side length $0.005$. At $t=40$%
, the vertical rectangle is seen to have become less excited than initially,
while the small square $E$ has combined with the central part of $B_\alpha$
and is becoming more excited.
At time $t=102.5$, $B_\alpha$ has become refractory, and the original excited region $E$ has
grown, with its outer ring excited but a yellow hole appearing in its center.

The refractory region seems to  
play a crucial role in forming the
$C$ shaped wave. Noting that by $t=102.5$ this refractory region is well 
established, we see from Figures \ref{2spots:wave}(a) and \ref{2spots:ps}(b) that
refraction arises noticeably sooner in the symmetric model. See the caption
to Figure \ref{2spots:ps} for further discussion.

Figure \ref{sym:startup}(e) and \ref{sym:startup}(f) show the
growth of the C-shaped wave. Its ends
have begun to curl, a critical characteristic. In Figure \ref{CurledLips}(a), 
the outer ring has closed but,
because the ends have curled, the closure has caused a small
excited seed to split from the outer ring,
\ref{CurledLips}(b). This excited seed
will move toward the center of the ring and grow as the closed outer wave exits
the computational region, soon closely resembling the C shaped wave in Figure \ref{sym:startup}(e).
It will then expand as in Figure \ref{sym:startup}(f), with curled ends
as in Figure \ref{CurledLips} and the cycle continues.
See https://pitt.box.com/s/ehu6dgm5citma3sniyrsb3twdmy9qsr3
At this point https://pitt.box.com/s/3pw98bpa51s9rvyynegk43exsk4higjs is also of interest, as it shows how our initial example, with a completely 
different initial pattern,  results in essentially the same fully developed pattern as in the example of this section, 
except for a 
rotation. 

J. Weimar \cite{Weimar} used a sophisticated cellular
automaton model to generate a series of slides showing a similar 
repetitive process. Another similar process, just on a smaller field, is
shown in \cite{ZykovBodenschatz}. In each of these papers the waves shown
started fully fledged. Our work shows that the same sort of repetition
can arise from a pattern which contains only cells which are either excited
or at equilibrium. We are including patterns like those shown by others 
because previous researchers were studying patterns constructed to show
steady rotations, while we must demonstrate that our patterns approach the
same cycles as theirs.

\begin{figure}[]
\begin{center}
\includegraphics[width=3in,height=1in]{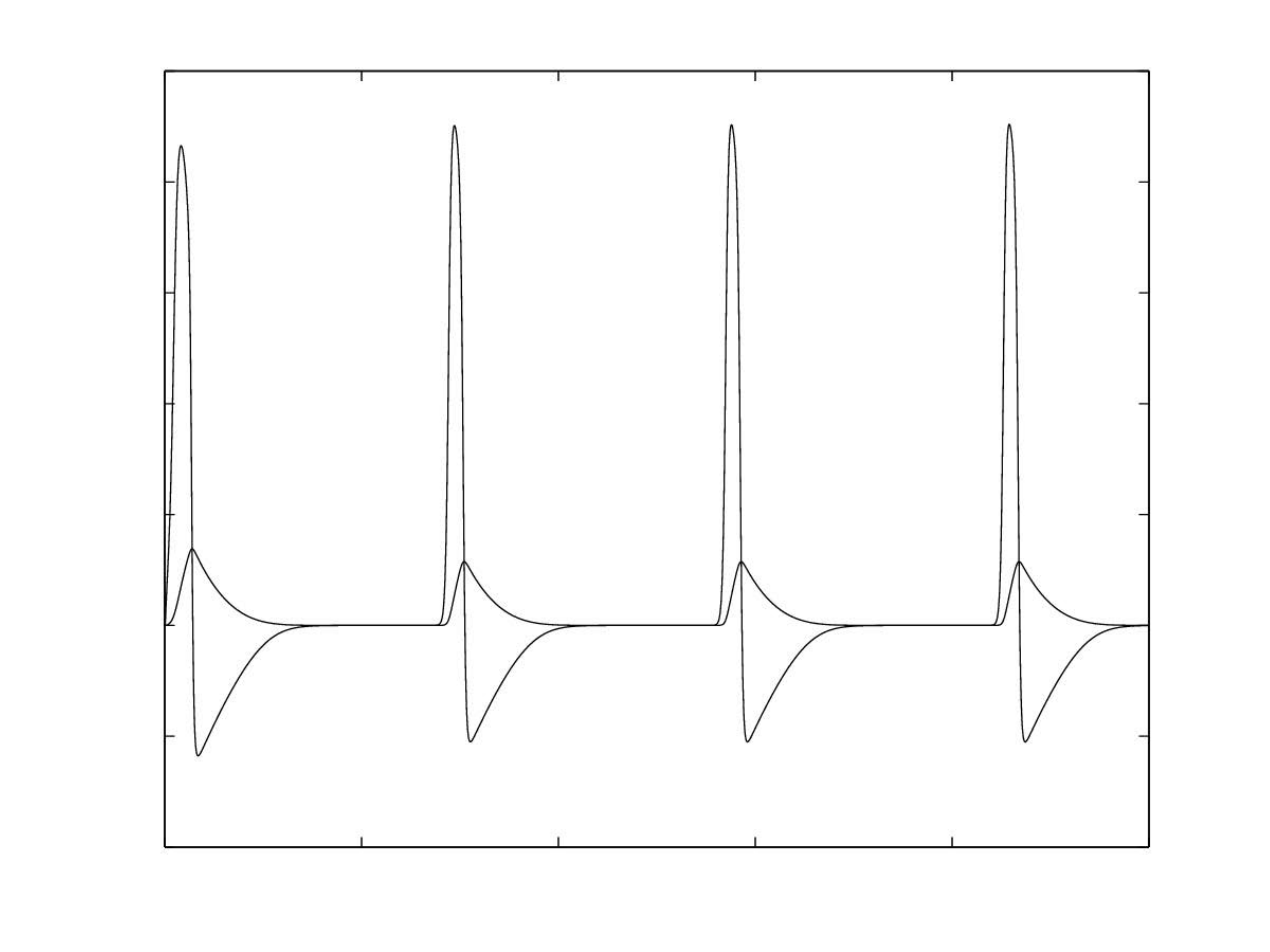}
\end{center}
\caption{Time history of $u$ and $v$ at the center point of the
calculational field for several repetitions. $u$ is the curve with higher
peaks and lower valleys, $v$ the one with smaller peaks and no valleys.
Visual periodicity is apparent in the fully developed cycles. }
\label{sym:timeHistory}
\end{figure}

Figure \ref{sym:timeHistory} shows the history of $u$ and $v$ at the center
point of the computational field. This point is $0.02$ to the left of the
initial excited region $E$, so it begins at equilibrium. The visual 
repetition of $u$ and $v$ in the final three cycles shown supports the opinion that
the solution is approaching periodicity.

\section{A more complex model\label{RandomCase}}

In our early search for repetitive behavior, we used pseudo-randomly generated
and positioned rectangles in a field otherwise at equilibrium.
 For each case of this type our pseudo-random number
generator produced a set of randomly placed ``spots'', which were
rectangular sets of cells in the mesh with dimensions $m\times n$, meaning $%
m $ cells along a horizontal edge and $n$ cells along a vertical edge. Here $%
m$ and $n$ were chosen randomly in a specific range for each spot. These
spots were placed well into the interior of the square $\Omega $.  We set $%
u\left( x,y,0\right) =0.8$ on each cell in these spots, $u=0$ elsewhere, and 
$v\left( x,y,0\right) =0$ everywhere. 

In these random simulations we used square
domains of between $101$ and $601$ mesh cells on a side.\footnote{Choosing an odd 
mesh size had a programming advantage in a cases where we later enlarged the field as
described below} After
experimentation we used spots which were between $3$ and $7$ cells on a
side. We continued a run until the maximum value of $u$ over the entire
field was less than $b$, since further excitation was then impossible.%
\footnote{This can be shown by phase plane analysis of (\ref{2.3},\ref{2.4}).}
 If this did
not happen within a predetermined maximum number of time steps, (often $4000$%
), then the starting configuration was saved for further study.

We extended the runs of successful trials to times well beyond the
initial screening and in all but a few cases the wave persisted over many
cycles around the region. The example we discuss here, like most,  appeared to approach
a periodic pattern. 
Before discussing it in detail, we wish to point
out that for this model, it was not particularly rare for the pattern to
survive to the point of apparent persistence. In a trial with $b=0.169$ of
1000 random choices of up to 35 spots in a 201 x 201 mesh, 37 of them lasted
4000 steps, or until $t=2000$. And we believe, based on some integrations
for hundreds of thousands of time steps, that all of the surviving patterns
in this case would have continued indefinitely.  We hope in the future to
examine this question for a model which bears a closer relation to a
particular physical setting. The article \cite{Tyson} is about one such
model.

\begin{figure}[]
\begin{center}
\begin{tabular}{cc}
\includegraphics[width=1.2in, height= 1.2 in]{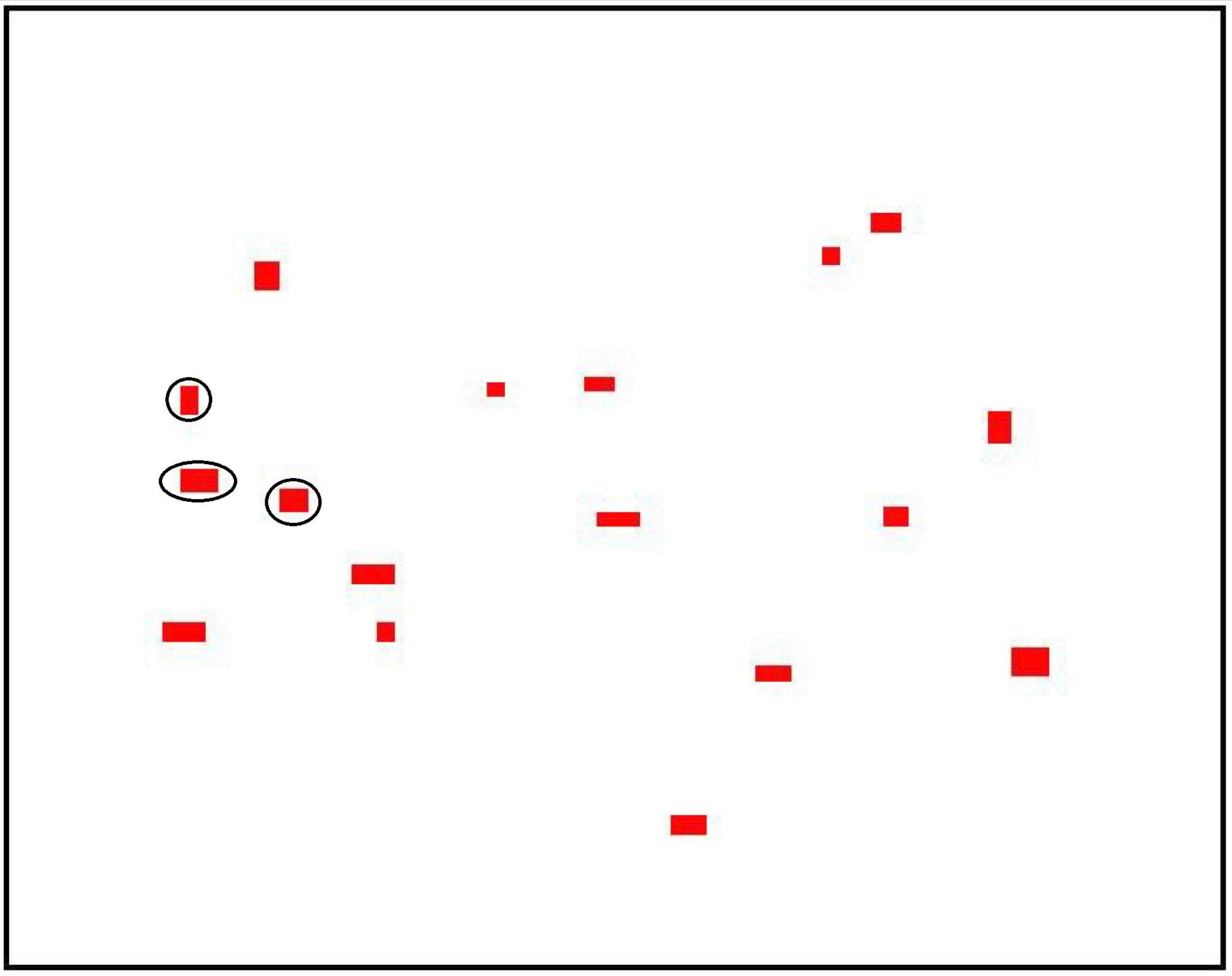} & 
\includegraphics[height= 1.2 in, width = 1.2 in]{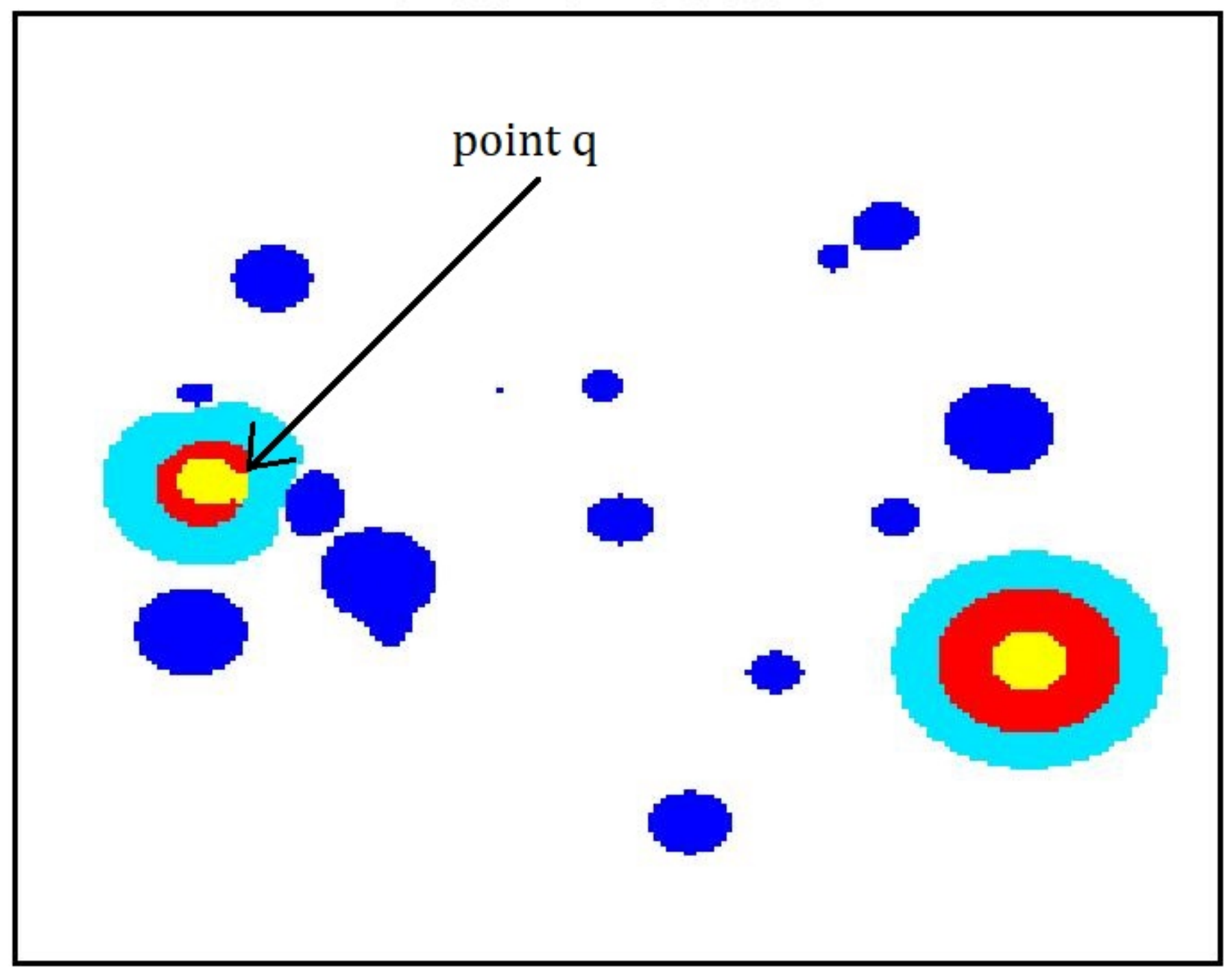}  \\ 
(a) $t=0$ & (b) $t = 125$  \\
\end{tabular}%
\end{center}
\caption{(a) A pseudo-randomly chosen set of rectangular spots for the initial stimulation, with 
$u=0.8$ in these spots and $u=0$ elsewhere. Everywhere,$v=0$ . If any one of the three circled spots 
is removed, then the remaining pair is subcritical. But all three work 
together to produce a $C$-shaped wave. (b) The target pattern on the right is
 a snapshot of an outgoing wave, with a blue wave front leading, followed by the red excited area and 
yellow wave back. On the left a similar, though indented, blue wave front is seen, but the excited ring 
has been broken leaving a red C shape. Meanwhile, all other spots of initial excitation have 
become refractory. Point $q$ was described in the discussion of Figure \ref{2spots:excited}. Here $b=0.169$. }
\label{random-startup-cde}
\end{figure}

We will go into less detail here than
we did for the two-spot example. Figure \ref{random-startup-cde} shows the 
initial condition and then jumps to a point where a C shape excited region and point
$q$ have appeared. Note the similarity between the C shaped excited regions on the left side
of Figure \ref{random-startup-cde}(b)
and in Figure \ref{2spots:excited}(d). \footnote{There are differences in the previous and subsequent
steps in
these two examples which may be of interest in future efforts to understand these processes better
but are not needed at this stage.} 

The excitation pattern on the right side of Figure \ref{random-startup-cde}(b)
is the beginning of an outgoing target wave spreading from a center, with
the dynamics dominating in the center and producing a reinvigorated red disk
inside the light blue ring. In this right target pattern, the
front and the back do not meet, that is, they do not share any boundary
points, and there is no point $q$.  Left to itself, it would expand beyond the domain shown
and not cause repetitive motion.

Any initially stimulated spot not involved in either the target pattern on 
the right (produced by one spot) or the C-shaped excited region on the left (produced by
the three circled spots in Figure \ref{random-startup-cde}(a) ), is subcritical.

In Figure \ref{random-startup-ghi}(a), each of the excited regions has
grown, so that they are now in contact, and it looks like the $C$
shape is going to take a bite out of the circular excited ring. This
explains why the circular wave on the ring on the right does not kill
off the activity on the left, as might have been expected.

In Figure \ref{random-startup-ghi}(b) we see that the C-shaped wave 
has cut the target pattern wave in two places, leaving a segment in the middle
which, it turns out, survives on its own. In Figure \ref{random-startup-ghi}(c)
we see that a new C-shaped wave has developed out of that segment, with a 
surrounding outgoing
wave. Further, the large surrounding wave front is, in
mathematical terms, topologically equivalent to an annulus. There
continue to be two points $q$, at the two tips of the inner $C$ shaped
wave.
In this situation,
we believe that what happens outside the annulus can have no effect on what
happens inside, because the trailing refractory region prevents stimulation
from the outside. For this to be possible in the models we are considering,
it appears necessary that the domain be a certain minimum size, or larger.
This opinion is supported below in the section on boundary conditions
and mesh adequacy.
\begin{figure}[tbp]
\begin{center}
\begin{tabular}{ccc}
\includegraphics[height=1. in, width =1. in]{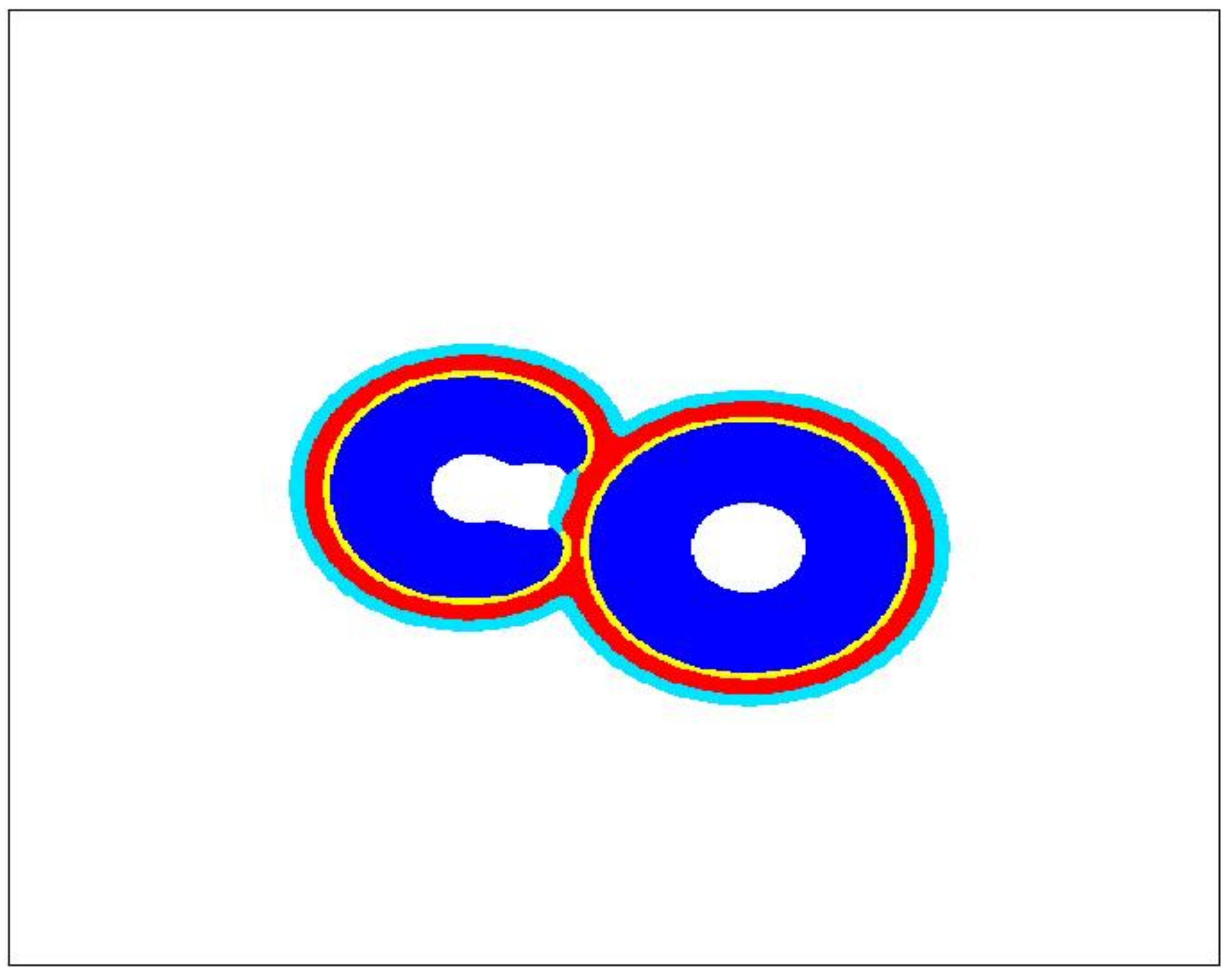} & %
\includegraphics[height=1. in, width =1. in]{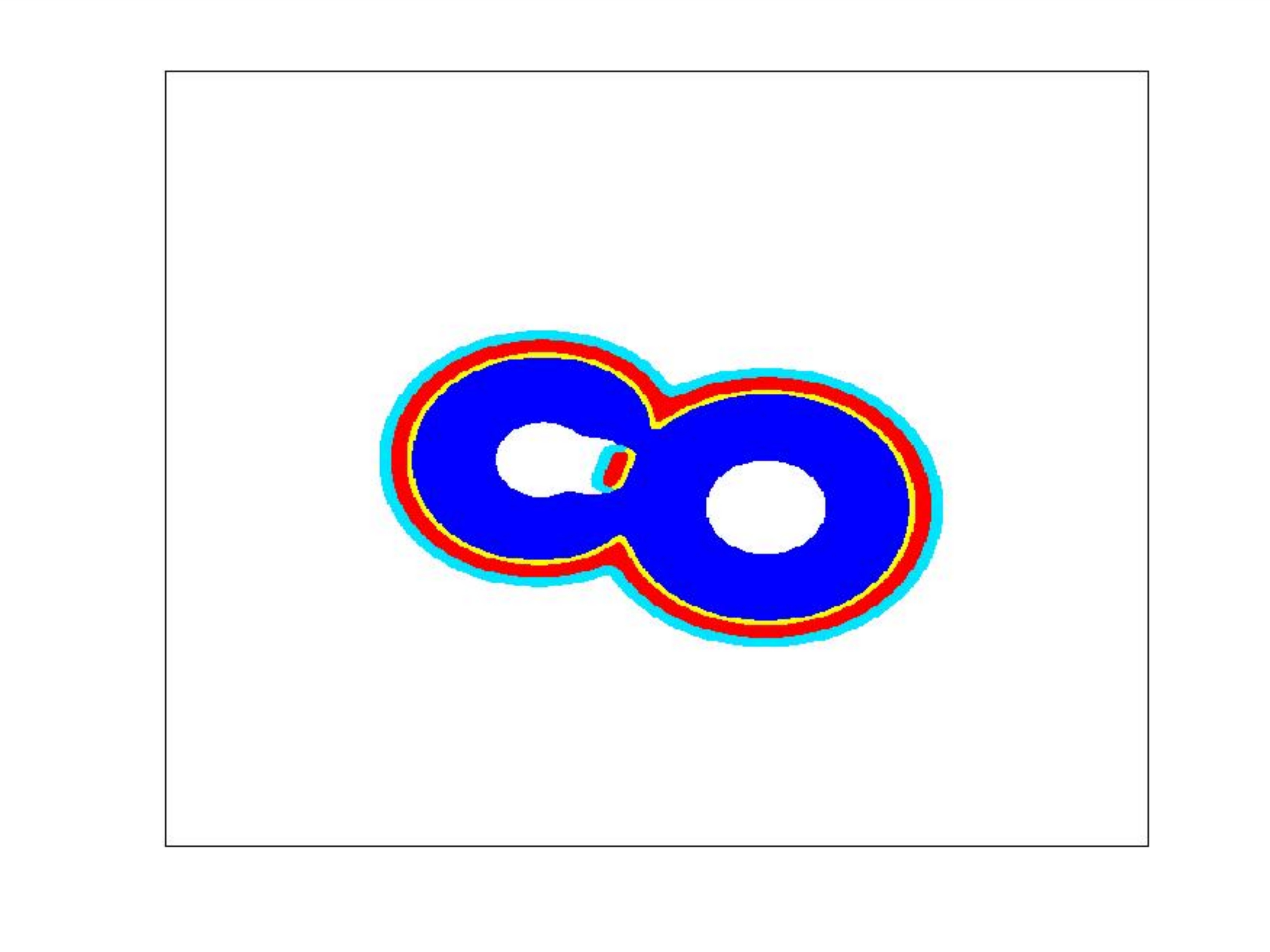} & %
\includegraphics[height=1. in, width =1. in]{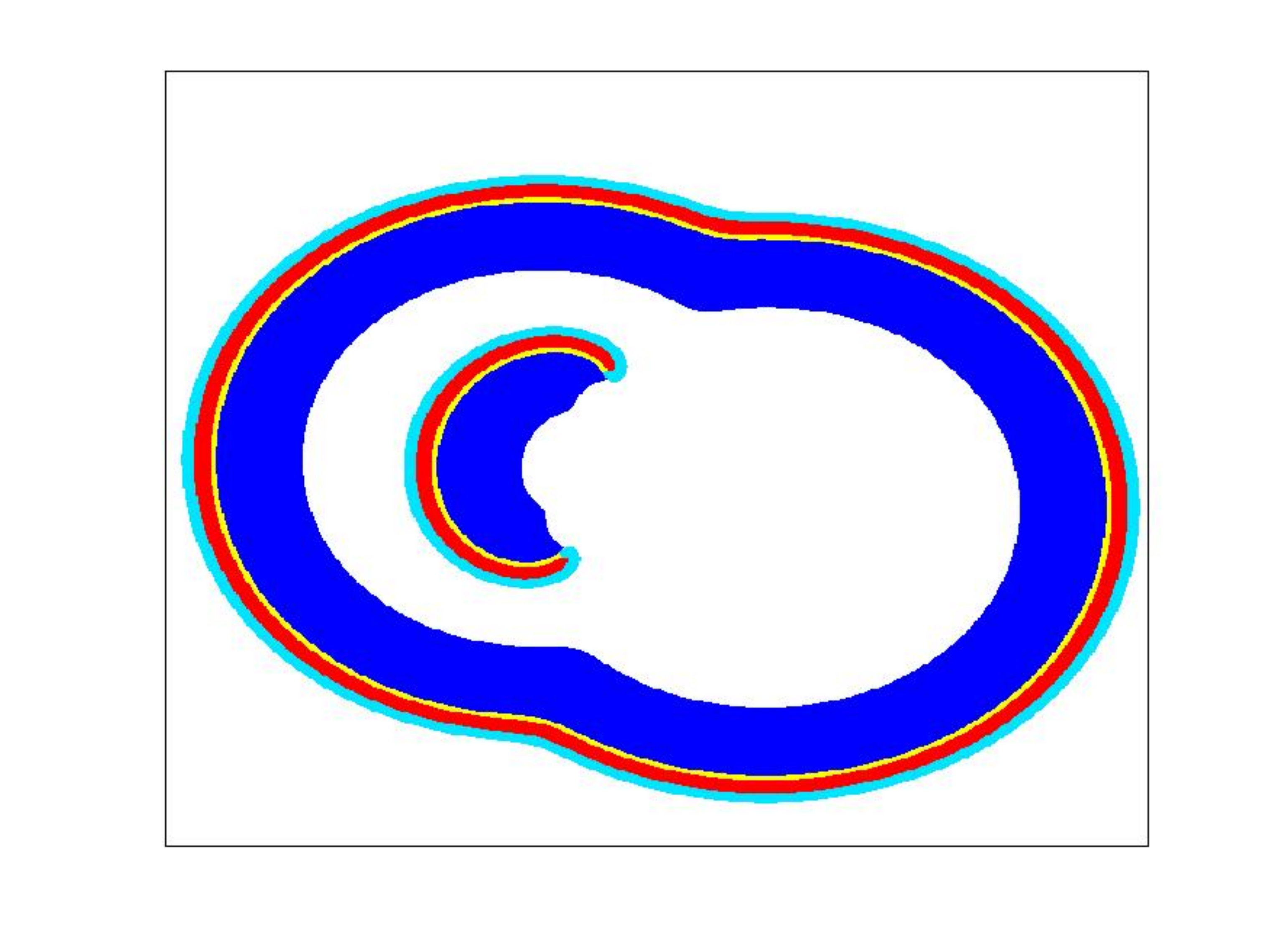} \\ 
(a) $t = 600$ & (b) $t=675$  & (c) $t=1325 $\\
&   &
\end{tabular}%
\end{center}
\caption{Continuation of Figure \protect\ref{random-startup-cde} to later
times. (a) The C-shaped wave on the left side of Figure \ref{random-startup-cde}(b)
has grown and moved to meet the growing target pattern on the right. It appears
to ``take a bite'' out of the approaching circular wave. (b) The interaction between
the two waves results in a new ``seed'' moving to the left. See https://pitt.box.com/s/5tlzjnq55etlnccof3mbybba3x72mps9.  (c) The pattern has settled
into apparent periodicity, with a time plot 
similar to that in Figure \ref{sym:timeHistory}.    
}
\label{random-startup-ghi}
\end{figure}

\section{Other examples}

\label{other examples}

\subsection{Opposing spirals}
\label{OpposingSpiralsExample}

\label{opposing}

\begin{figure}[h]
\begin{center}
\begin{tabular}{ccc}
\includegraphics[width=1.0in]{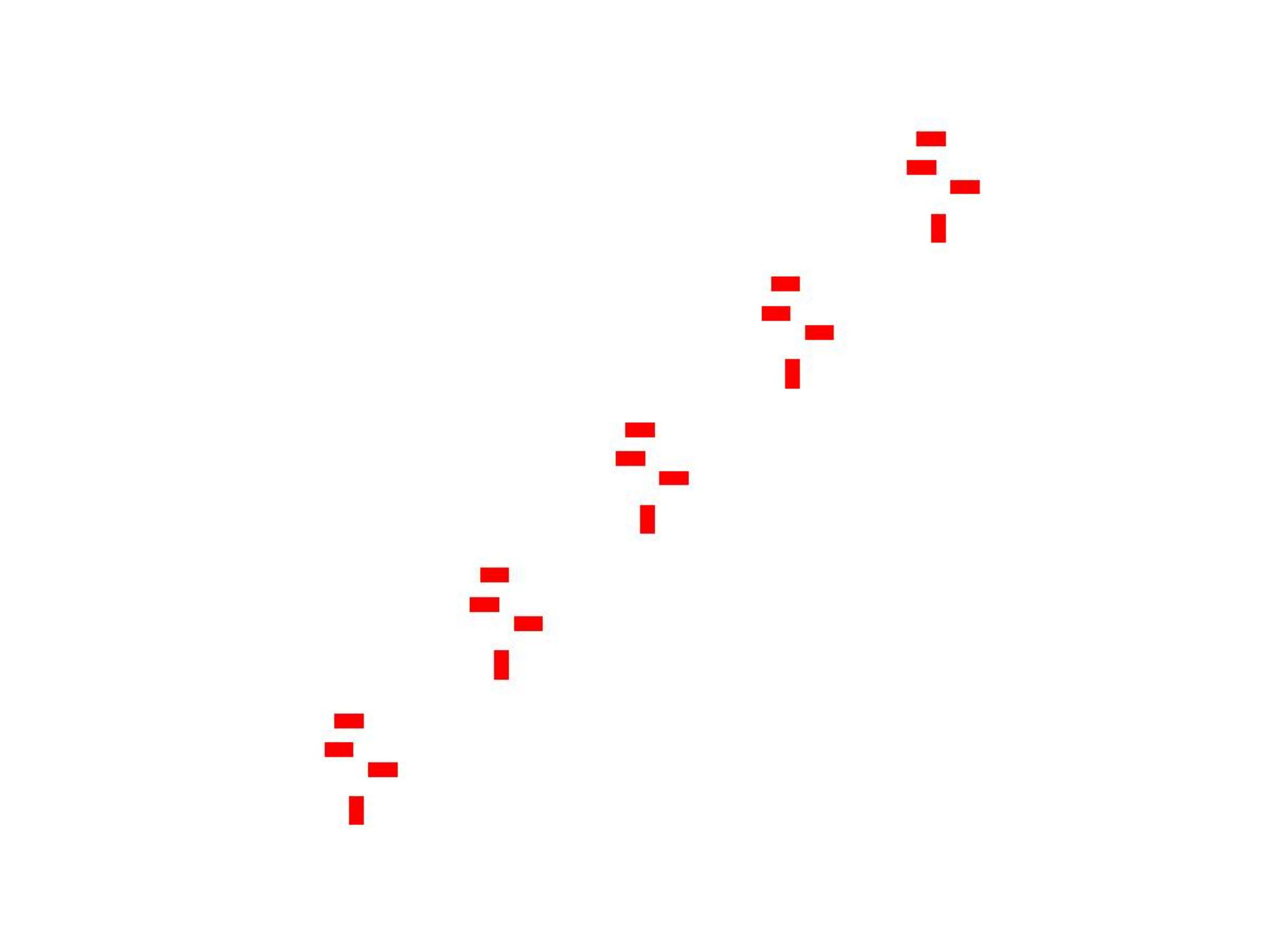} & %
\includegraphics[width=1.0in]{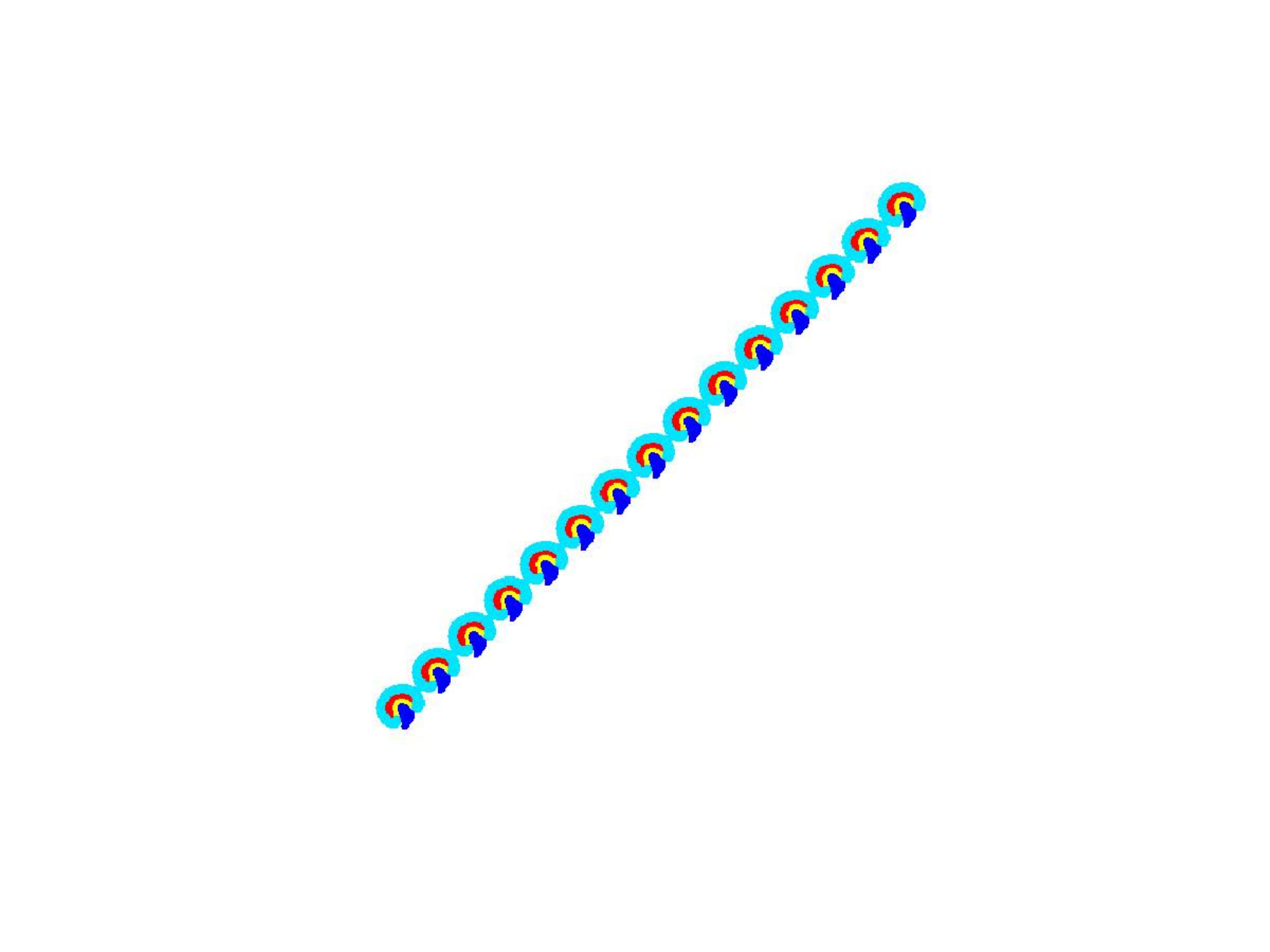} & %
\includegraphics[width=1.0in]{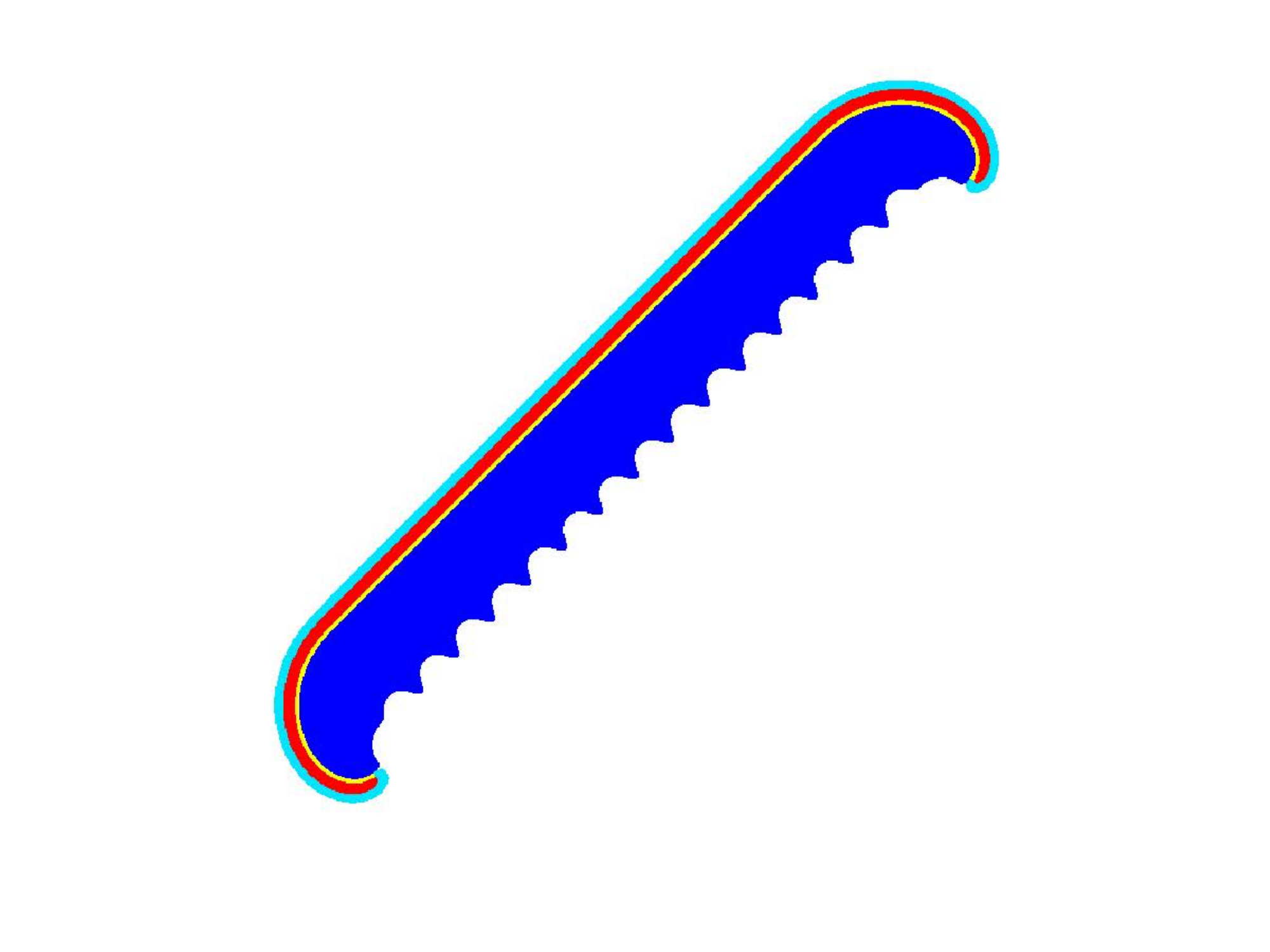} \\
(a) $t=0$ & (b) $t = 140$ & (c) $t = 650$
\end{tabular}\\
\includegraphics[width=1.4 in]{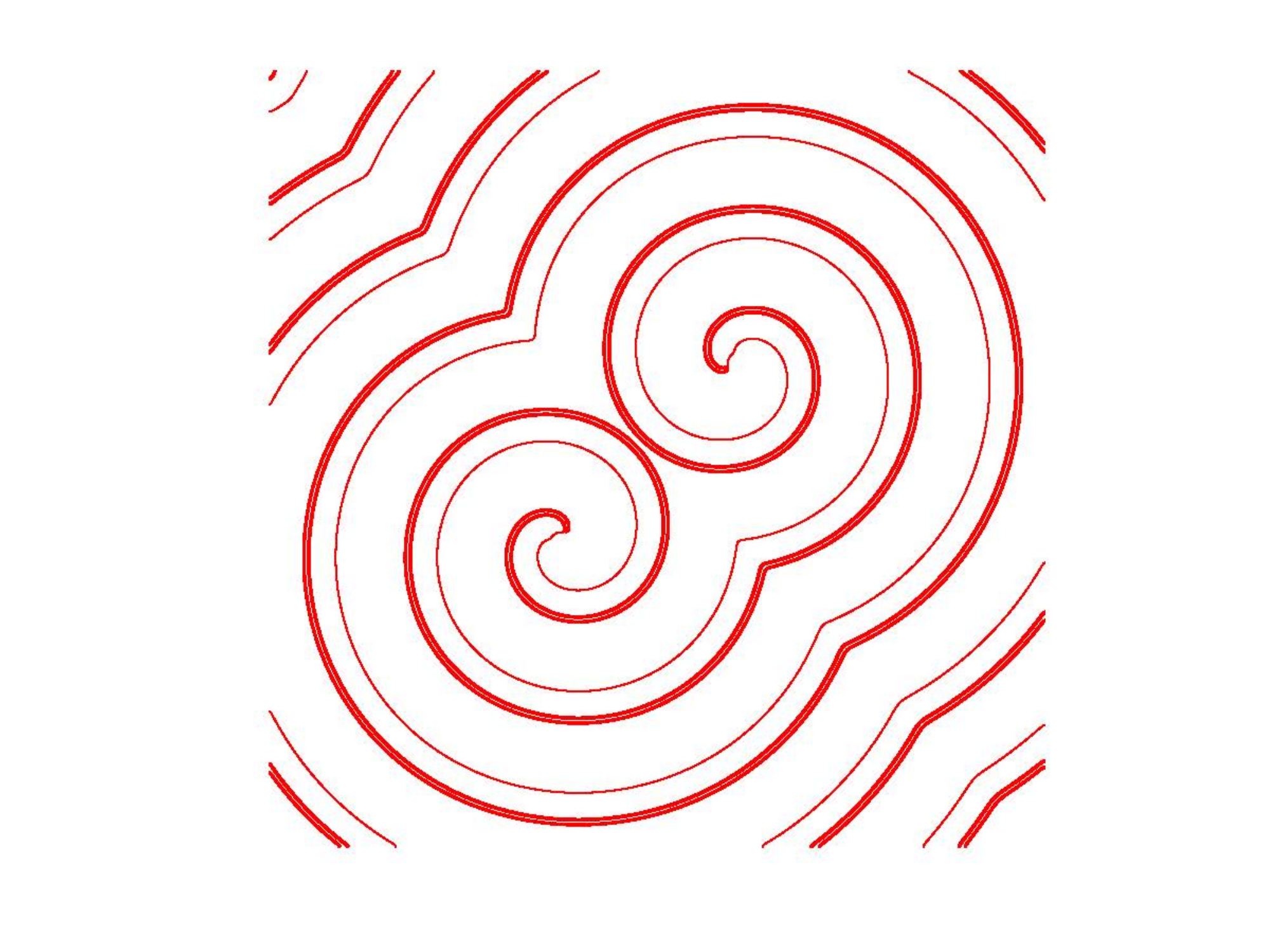} \\
(d) Fully developed
\end{center}
\caption{Two opposing spirals: (a) Shows a detail of the initial pattern;
(b) and (c) show early evolution; and, (d) shows excited contours of	
$u$ in the developed spiral pattern.
Scales in the four frames are not the same.}
\label{OpposingSpirals}
\end{figure}

We present an example of two opposing spirals surrounded by closed curves
expanding outward, as shown in Figure \ref{OpposingSpirals}. This
figure is a snapshot at problem time $t=10,000$ of a series of outgoing
waves whose central core consists of two spirals. This situation is often
observed in experiments, for example in Zykov \cite{Zykov1}, Figure 1.

In early random examples, a set of four rectangles was found to generate
a sequence similar to the one in Figure \ref%
{sym:startup}. Assembling several of these sets of rectangles in a line
was then found to separate the ends of the C-shaped wave sufficiently for
them to curl up into spirals. For this example, fifteen identical sets of
rectangles are arranged in a diagonal line as indicated in Figure \ref%
{OpposingSpirals}(a). Only five of the sets of rectangles are
shown, so that the details are clear.

In Figure \ref{OpposingSpirals}(b), showing time $t=140$, the individual
sets are separately developing and are all shown and in Figure \ref%
{OpposingSpirals}(c), showing time $t=650$, they have combined
into a long, narrow C-shaped wave oriented at an angle of $\pi/4$. The ends
of this C-shaped wave curl into spirals which don't touch until after a full
turn.  Figure \ref{OpposingSpirals}(d) shows the spiral configuration after
a long time.  This configuration is not steady, but repeats itself periodically.

\subsection{A two-arm spiral}

\label{two arms}

This example shows a remarkable tendency to approach a two armed spiral from
earlier stages where there is no sign of spiral behavior. Experimental
evidence for the existence of multi-armed spirals is shown in \cite{Muller4}. This
example suggests that two-armed spirals exist and have a large 
basin of attraction.  


\begin{figure}[]
\begin{center}
\begin{tabular}{cc}
\includegraphics[width=1.2in, height=1.2in]{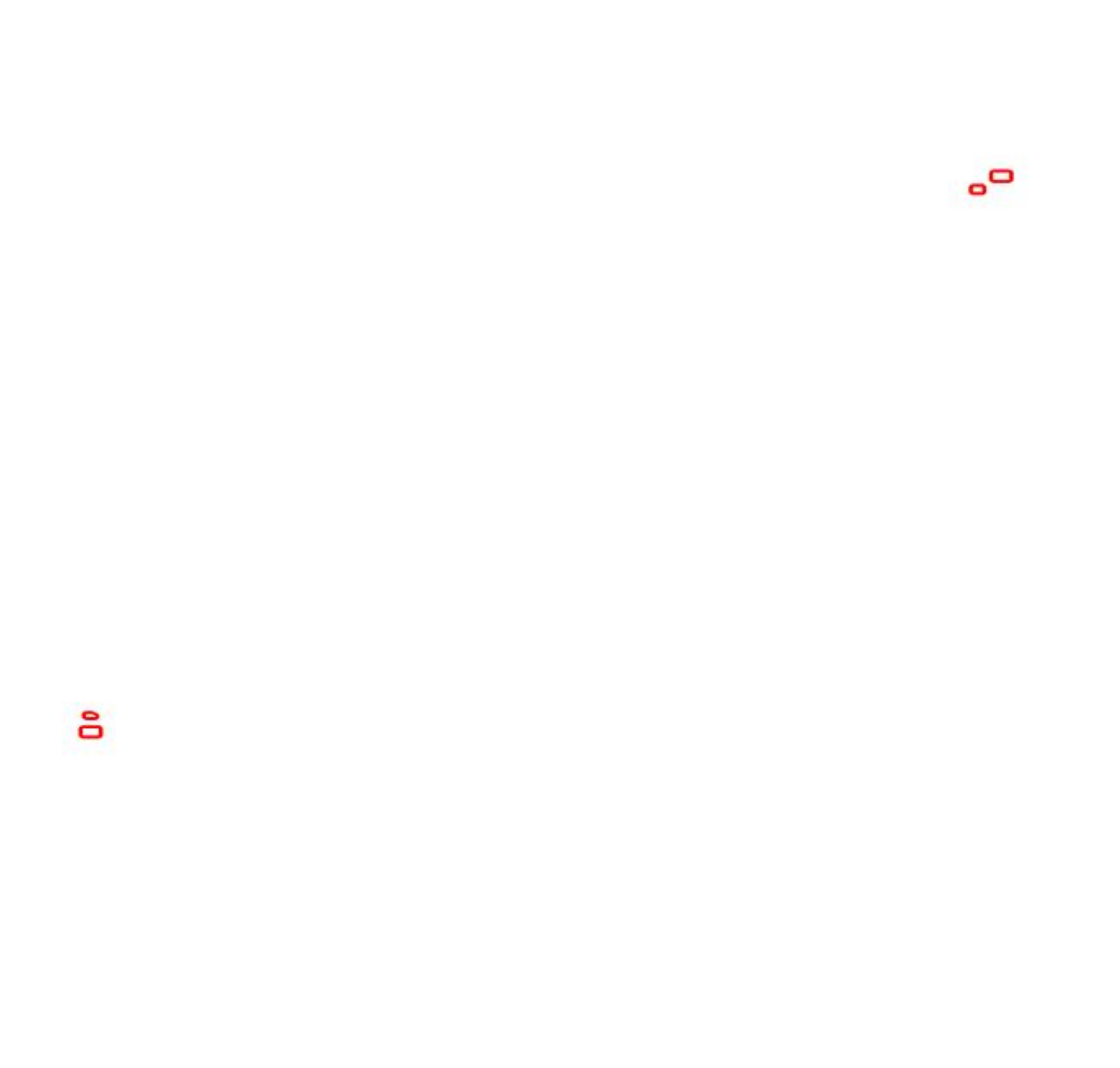}  &%
\includegraphics[width=1.2in,height=1.2in]{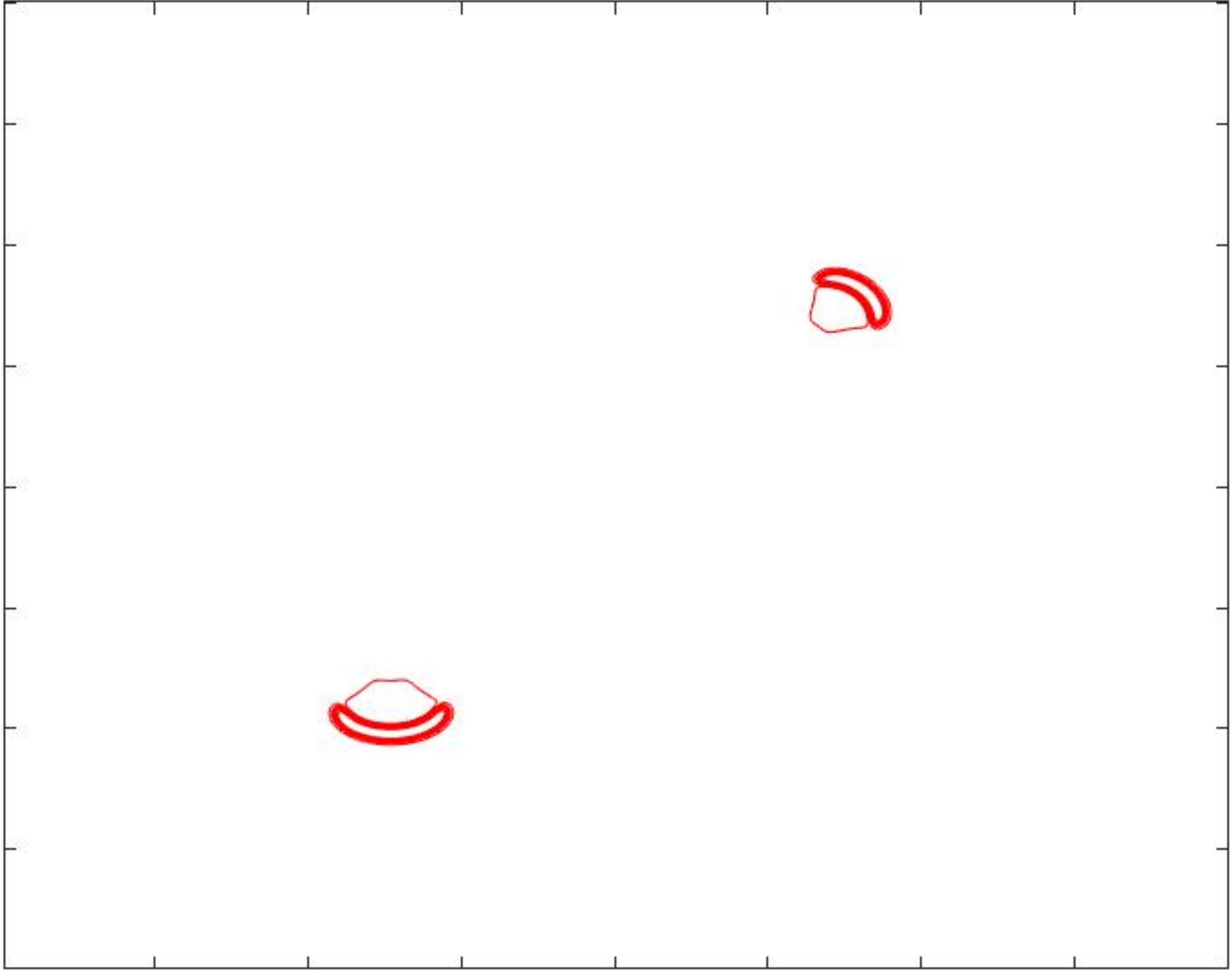}  \\
(a) $t=0$ enlarged & (b) $t=50$ \\
\includegraphics[width=1.2in,height=1.2 in]{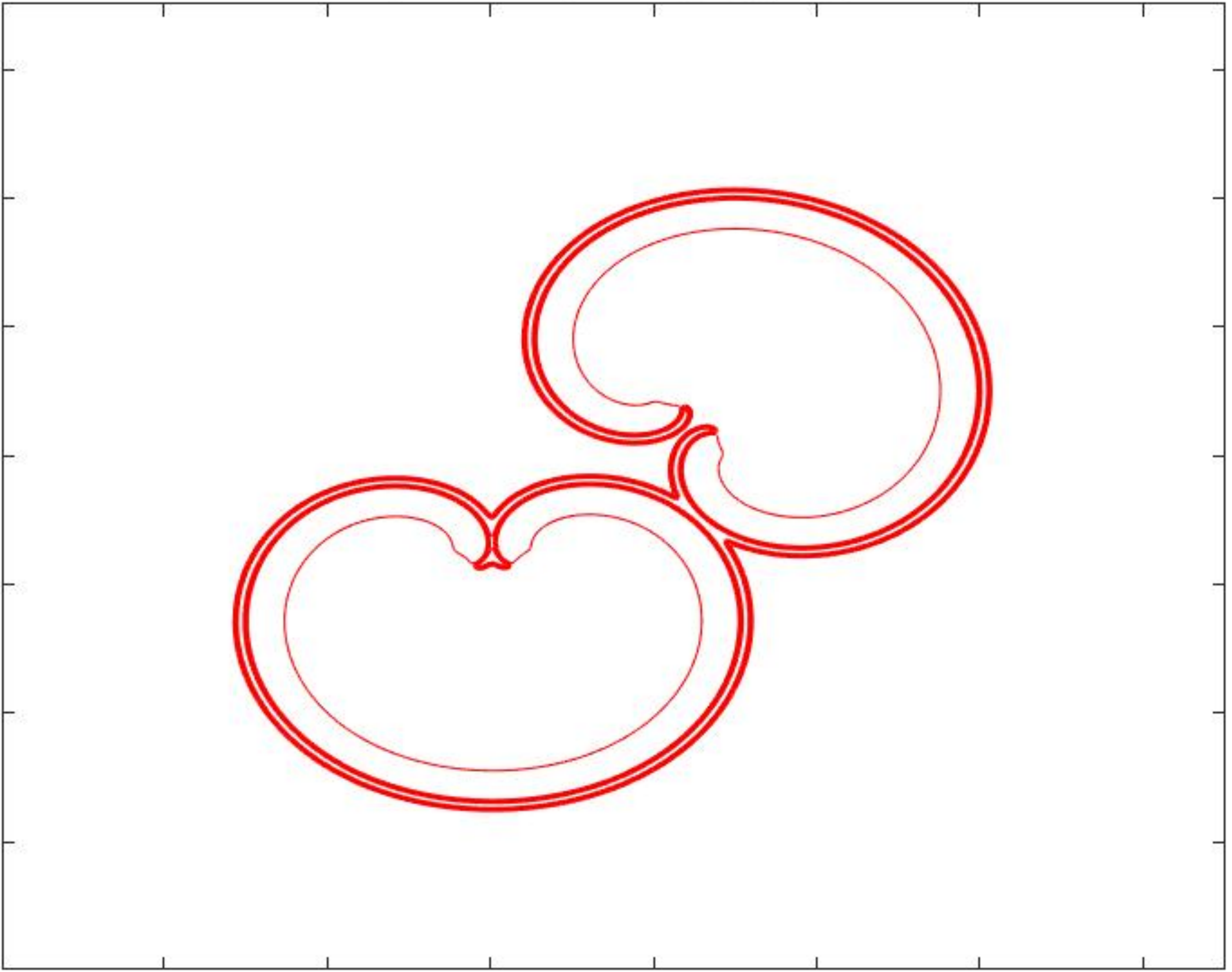}  & 
\includegraphics[width=1.2in,height=1.2in]{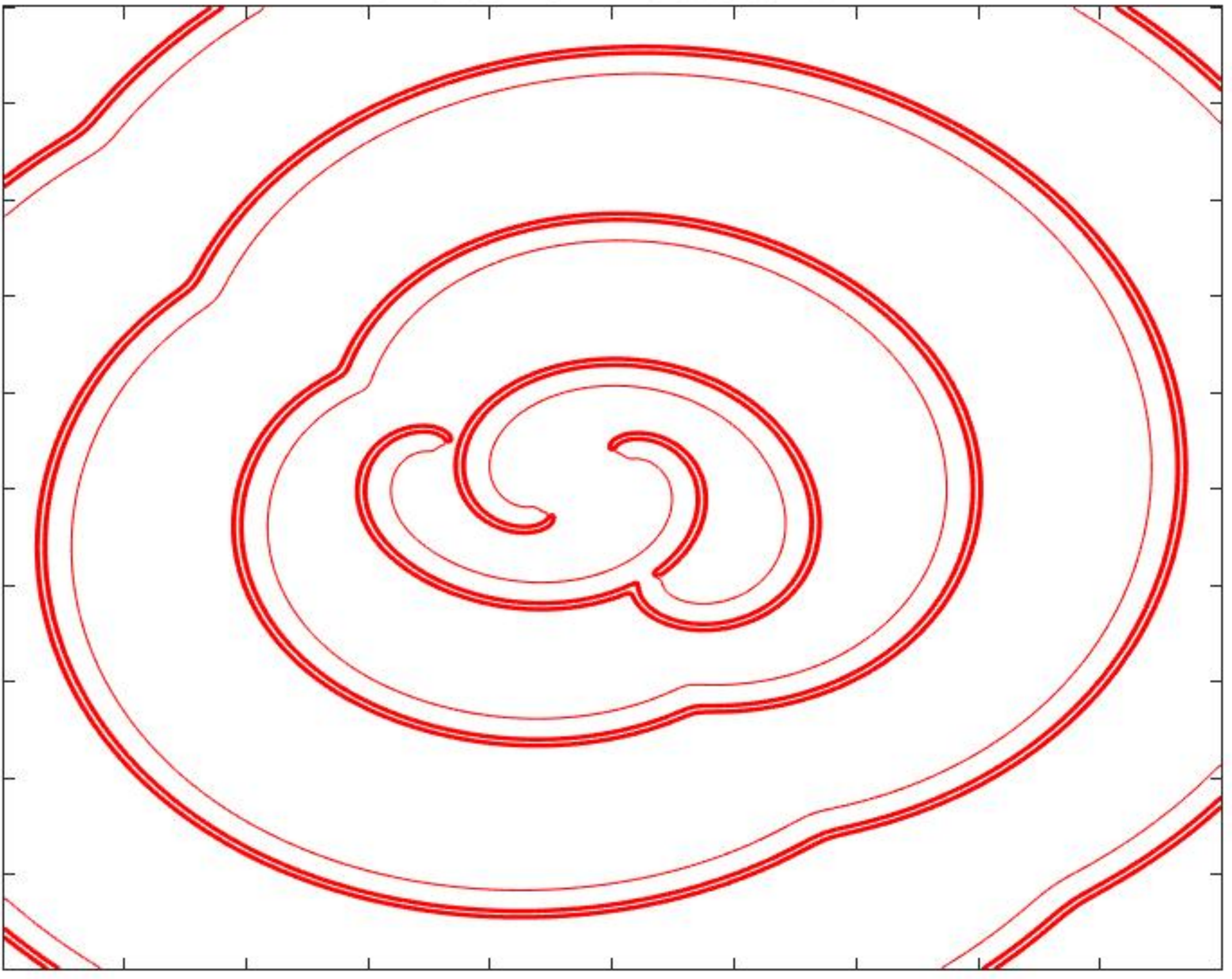} \\
c) $t=2148.5$  & (d) $t =17700$ \\
\includegraphics[width=1.2in, height=1.2 in]{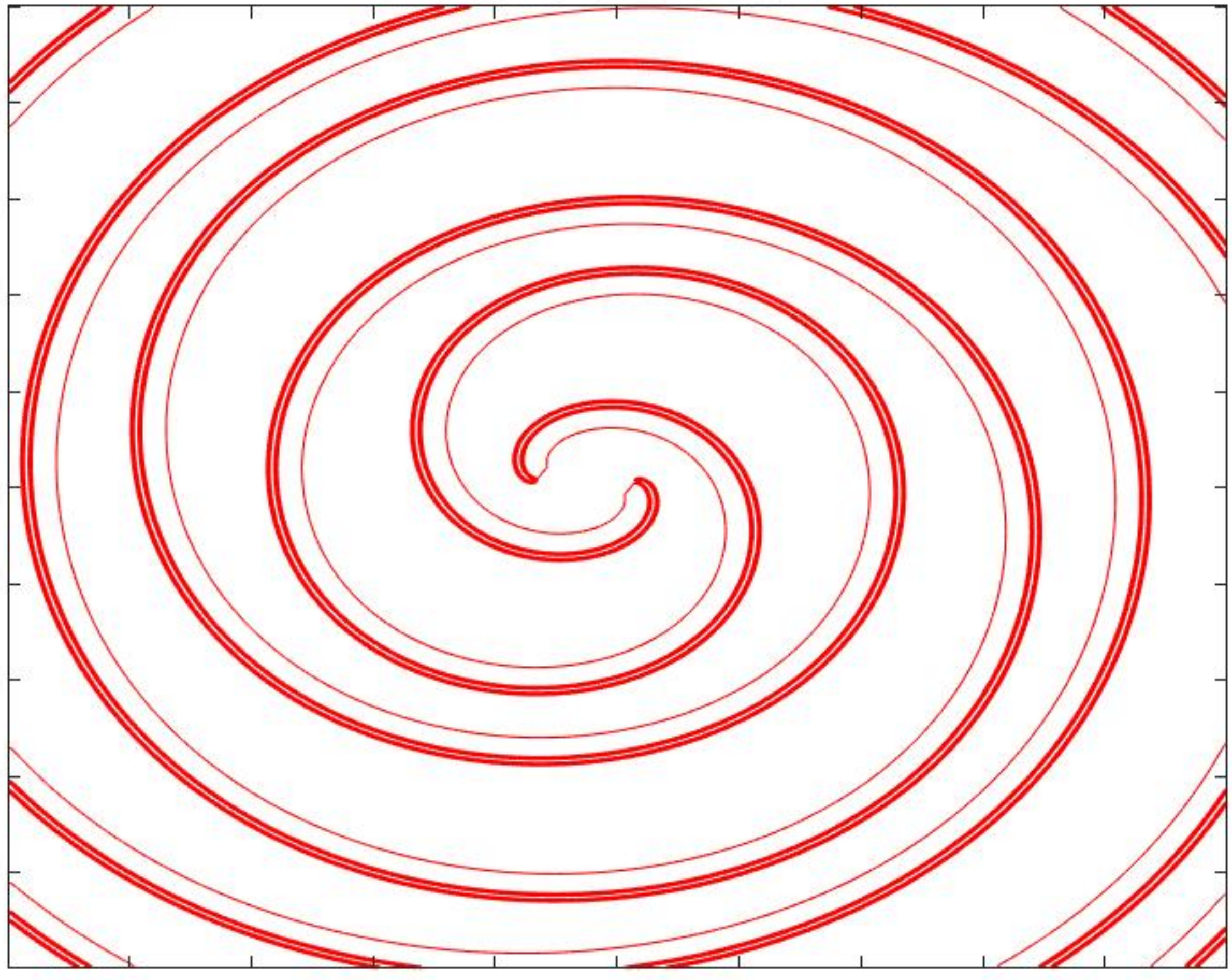}  \\ 
 (e) $t=65550$ \\
&
\end{tabular}%
\end{center}
\caption{ (a)The start position; note two pairs of spots. (b), (c), 
(d) Three intermediate patterns, (e) A final rotating
two-armed spiral.  The scales changes, with the distance between
the two pairs of spots in (a) roughly equal to the distance between
the two central spiral tips in (e). 
Here $b=0.1765$. See https://pitt.box.com/s/j7uqz6px4jm1g0sc7fbr6u60rr2j4wp5.
}
\label{doublespiral}
\end{figure}

\section{Numerical Methods}

\label{discretization}

The results in this paper are all from computational experiments performed using files written for a Matlab computer package.  These files are available upon request to the authors.   The numerical methods used in the program are summarized in the following paragraphs. 

The discrete form of (\ref{2.1}-\ref{2.2}) chosen for this study involves
the standard five-point discretization of the Laplacian on a uniform, square
spatial mesh. Away from boundaries, this discretization introduces a
truncation error proportional to $\Delta x^2$, where $\Delta x$ represents
mesh spacing. Neumann conditions at the boundary are implemented by
reflection, introducing an error proportional to $\Delta x$. As discussed
below, boundary errors are small and localized near the boundary itself and
do not affect conclusions reached in this paper.

An explicit Euler discretization with uniform step size is used for the
temporal term. This approach introduces an error proportional to $\Delta t$,
the step size. It also requires a limit on time step size in order to
maintain numerical stability. For most of the simulations used for this
paper, these limitations are not onerous.

\subsubsection{Boundary conditions and mesh adequacy}

\label{boundary conditions}

In order to confirm that boundary conditions do not have a significant
effect on the simulations in this paper, two simulations are carefully
examined. The first is the evolution of a seed from the initial
configuration with no refractory regions into an arc-shaped wave front that
closes and splits into an outwardly-moving closed wave front and another seed
in its interior (as in Figures \ref{sym:startup} and \ref{CurledLips}).

Numerical experiments on square computational regions of between 201 and 601
mesh blocks of $\Delta x= 5.0\cdot10^{-3}$ along each side show that the
initial evolution of the seed takes place far from boundaries and is
independent of boundary placement. The evolution of the arc-shaped wave front
to its closure is not affected by boundary placement so long as the
wave front does not get within about $5\Delta x$ of the boundary. The newly
created seed remains in the interior, away from the boundary, and the
outwardly-moving wave front eventually passes out through the computational
boundary without influencing the interior.

A second numerical experiment compared the evolution of a seed configuration
into the periodic double spiral in Figure \ref{OpposingSpirals}(d)
using two different boundary conditions:
Neumann and Dirichlet. The computational region is a $10\times10$ square of $%
2000^2$ square mesh blocks the same size as above. The relative Euclidean
norm of the difference between the two solutions computed on an inner region
of size $9.75\times9.75$ at the same time is $8.6\cdot10^{-5}$, and still
less than 5\% on an inner region of size $9.875\times9.875$.

Most simulations in this paper were performed using a mesh increment $\Delta
x=5.0\cdot 10^{-3}$ and time step size $\Delta t=0.5$. To determine that
this selection is adequate to simulate solutions of (\ref{2.1}-\ref{2.2}),
a Richardson type error analysis was performed \cite{Atkinson}. A sequence
of one-dimensional simulations of an advancing wave front were completed
using progressively finer sizings. Each simulation used half the $\Delta x$
and a quarter $\Delta t$ of the preceding simulation. The sequence
exhibited convergence at only slightly larger than the expected $0.25$ rate,
justifying the original choice as within the asymptotic range.

\section{Conclusions}

   We have presented very strong numerical evidence that the FitzHugh-Nagumo model of an excitable medium exhibits spiral waves and other recurring 
patterns in a spatially homogeneous simply connected region starting from initial conditions consisting only of excited regions and regions at 
equilibrium.  
We found that in this model these patterns can occur relatively frequently with random generation of rectangular spots of excitation. 

We developed enough understanding of why this happens to be able to construct special minimal sets of spots which acted as ``seeds'' for such behavior, 
including some with only two spots. Always these seeds included at least one spot which, if placed in a large field otherwise at rest, 
would die out without creating an outgoing wave. In some cases, all the spots involved had this property. 

We cited references which present strong arguments for the importance of these waves in serious clinical conditions, including ventricular fibrillation, 
depression, and epilepsy. Further, our study with random initial conditions leads us to suggest that these patterns may sometimes occur
``spontaneously", from random input. We believe that experimental testing of our results is highly desirable.

\medskip

\begin{center}
ACKNOWLEDGEMENTS
\end{center}

\medskip

We wish to thank G.~B.~Ermentrout, W.C. Troy, J.~Tyson, and J.~Weimar for encouragement to pursue this problem, and J.~Weimar  
for providing software which he had used for the FitzHugh-Nagumo equations and which contained a module which allowed for randomization with excited and refractory cells. 
While we eventually used different software, it was randomization, using only excited cells, which led to success once proper parameters had been found.    

We also wish to thank the Center for Research Computing at the University of Pittsburgh for the use of its facilities during this project, and R. Roskies for 
suggesting and facilitating this assistance.

\appendix

\end{document}